\documentclass[11pt]{article}
\usepackage{a4,epsfig}

\usepackage[round,colon]{natbib}
\renewcommand{\cite}{\citep}
\newcommand{\bfig}[3]{\begin{figure}\vspace*{#2}\begin{center}%
\leavevmode\epsfxsize #1\epsfbox{figures/#3.eps}}
\newcommand{\efig}[2]{\end{center}\vspace*{-0.2cm}\caption{#2%
\label{fig:#1}}\end{figure}}

\newcommand{\be}{\begin{equation}}
\newcommand{\ee}{\end{equation}}
\newcommand{\bea}{\begin{eqnarray}}
\newcommand{\eea}{\end{eqnarray}}
\newcommand{\lav}{\left\langle}
\newcommand{\rav}{\right\rangle}
\newcommand{\gdot}{\dot{\gamma}}
\newcommand{\la}{\lambda}
\newcommand{\de}{\Delta}

\newcommand{\D}{\displaystyle}

\def\(#1){(\ref{#1})}

\begin{document}

\title{Ageing and Rheology in Soft Materials}

\author{S. M. Fielding\\Department of Physics and Astronomy, University of
Edinburgh,\\JCMB,
King's Buildings,\\ Mayfield Road, Edinburgh, EH9 3JZ, UK.
\and P. Sollich\\Department of Mathematics, King's College,\\ University of
London, Strand, London,
WC2R 2LS, UK.
\and M. E. Cates\\Department of Physics and Astronomy, University of
Edinburgh,\\JCMB, King's
Buildings,\\ Mayfield Road, Edinburgh, EH9 3JZ, UK.}

\date{29 June 1999}

\maketitle
\begin{abstract}

We study theoretically the role of ageing in the rheology of soft
materials.  We define several generalized 
rheological response functions suited to
ageing samples (in which time translation invariance
is lost). These are then used to
study ageing effects within a simple scalar model (the ``soft glassy
rheology" or SGR model) whose constitutive equations relate
shear stress to shear strain among
a set of elastic elements, with distributed yield
thresholds, undergoing activated dynamics governed by
a ``noise temperature", $x$. (Between yields, each
element follows affinely the applied shear.) For $1<x<2$ there
is a power-law fluid regime in which transients occur,
but no ageing. For $x<1$, the model has a macroscopic yield stress. 
So long as this yield stress
is not exceeded, ageing occurs, with a sample's apparent
relaxation time being of order its own age. 
The (age-dependent) linear viscoelastic loss
modulus $G''(\omega,t)$ rises as frequency is {\em lowered}, but
falls with age $t$, so as to always remain less than
$G'(\omega,t)$ (which is nearly constant).  Significant ageing is
also predicted for the stress overshoot in nonlinear shear startup
and for the creep compliance. Though obviously
oversimplified, the SGR model may provide a valuable paradigm for
the experimental and theoretical study of rheological ageing
phenomena in soft solids. 

\end{abstract}

\section{Introduction}

\label{sec:intro}

Many soft materials, such as foams, dense emulsions, pastes and
slurries, display intriguing features in their low frequency shear
rheology. In oscillatory shear, for example, their viscoelastic
storage and loss moduli, $G'(\omega)$ and $G''(\omega)$, are often
weak power laws of shear
frequency~\cite{MacMarSmeZha94,KetPruGra88,KhaSchneArm88,%
MasBibWei95,PanRouVuiLuCat96,HofRau93,MasWei95},
while their nonlinear stress response $\sigma$ to shear strain of
constant rate $\dot{\gamma}$ is often fit to the form $\sigma=A +
B\dot{\gamma}^n$ (known as the
Herschel-Bulkley equation, or when
$A=0$, the power-law
fluid)~\cite{Holdsworth93,Dickinson92,BarHutWal89}.
The fact that such a broad family of soft materials exhibits similar
rheological anomalies is
suggestive of a common cause, and it has been argued that these anomalies
are symptomatic of the
generic presence in such materials of slow, glassy
dynamics~\cite{SolLeqHebCat97,long_El}. Indeed,
all the above materials share features of structural disorder and
metastability: large energy
barriers impede reorganization into states of lower free energy because
this would require
rearrangement of  local structural units, such as the droplets  in a dense
emulsion.  The term
``soft glassy materials'' (SGM's) has been proposed to describe such
materials~\cite{SolLeqHebCat97,long_El}.

Glassy dynamics are often studied using hopping (trap) models, in which
single particle degrees of
freedom hop by an activated dynamics, in an uncorrelated manner, through a
random free energy
landscape~\cite{Bouchaud92,MonBou96}. By incorporating strain degrees of
freedom into such a
description, Sollich and coworkers~\cite{SolLeqHebCat97,long_El} proposed
a minimal model, called
the ``soft glassy rheology'' (SGR) model, which appears to capture several
of the rheological
properties of SGM's, although (for simplicity) all the tensorial aspects of
viscoelasticity are
discarded. The model exhibits various regimes depending on a parameter
$x$ (discussed in more detail below) representing the  ``effective
temperature" for the hopping
process. When this is small ($x \le 1$) the model exhibits a glass phase
which shows some
interesting properties above and beyond the power-law anomalies in
viscoelasticity mentioned above.
Specifically, the model shows {\em ageing behaviour}: its properties depend
on the elapsed time
since a sample was prepared. This is because the population of traps
visited never achieves a
steady state; as time goes by, deeper and deeper traps dominate the
behaviour (a phenomenon known
as ``weak ergodicity breaking").  Broadly speaking, the system behaves as
though its longest
relaxation time is of order its own age.

\begin{figure}[h]
\begin{center}
\epsfig{file=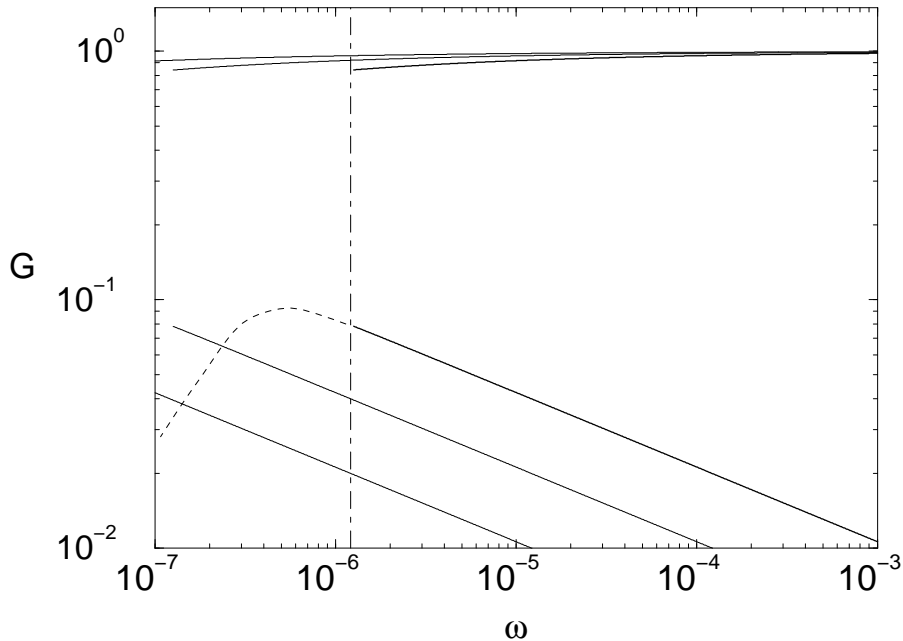,width=12cm}
\end{center}
\caption{Sketch of ageing scenario for dynamic moduli $G'$ (top) and
$G''$ (bottom). Assuming that data shown by the thick solid lines has
been measured, the conventional interpretation would be that $G''$ has
a loss peak at low frequencies $\omega$ (dashed) outside the
experimentally accessible range. But the rise of $G''$ towards small
$\omega$ could also be due to {\em ageing}: If the experiment is
repeated at later times, the thin solid lines (calculated from the SGR
model at noise temperature $x=0.7$; see
Sec.~\ref{sec:results_strain_linear_osc}) might be measured; the
putative loss peak would then always be unobservable because no
measurement is possible at frequencies below the inverse age of the
system.
\label{fig:possible_modes} } 
\end{figure}

The success of the SGR model in accounting for some of the generic
flow properties of SGM's suggests that a detailed investigation of its
ageing behaviour, and the effect this has on rheology, is now
worthwhile. Ageing has been intensively studied in the context of spin
glasses~\cite{BouDea95,CugKur95,BouCugKurMez98}, although some of the
earliest experimental investigations of it involved rheological studies
of glassy polymers~\cite{Struik78}. But we know of no previous theoretical
work that explores the link between ageing phenomena and
rheological properties within an explicit constitutive model. 
A particular added motivation is that detailed
experiments on rheological ageing, in a dense microgel suspension, are
now underway~\cite{microgel_beads}. Although various kinds of ageing
effects are often observable experimentally in soft materials, they
have rarely been reported in detail.  Instead they tend to be regarded
as unwanted obstacles to observing the ``real" behaviour of the
system, and not in themselves worthy of study. But this may be
illusory: ageing, when present, can form an integral part of a
sample's rheological response. For example, the literature contains
many reports of viscoelastic spectra in which the loss modulus
$G''(\omega)$, while remaining less than the (almost constant) storage
modulus $G'(\omega)$ in a measured frequency window, appears to be
increasing as frequency is lowered (see
Fig.~\ref{fig:possible_modes}). The usual
explanation~\cite{KosMorBat99} is that some unspecified relaxation
process is occurring at a lower frequency still, giving a loss peak
(dashed), whose true nature could be elucidated if only the frequency
window was extended.  This may often be the case, but an alternative
explanation, based on our explicit calculations for the SGR model, is
shown by the thin solid lines. No oscillatory measurement can probe a
frequency far below the reciprocal of the sample's age; yet in ageing
materials, it is the age itself which sets the relaxation time of
whatever slow relaxations are present.  Accordingly, the putative loss
``peak" can never be observed and is, in fact, a complete figment of
the imagination. Instead, a rising curve in $G''(\omega)$ at low
frequencies will {\em always} be seen, but with an amplitude that
decreases as the system gets older (typically ensuring that
$G''(\omega)$ never exceeds $G'(\omega)$).  Of course, we do not argue
that all published spectra resembling those of
Fig.~\ref{fig:possible_modes} should be interpreted in this way; but
we believe that many should be. The widespread reluctance to acknowledge
the role of ageing effects in much of the rheological literature suggests
that a full discussion of these could now be valuable. An exception
has been in the literature on ageing in 
polymeric glasses, especially the monograph by~\citet{Struik78}:
we return shortly to a brief comparison
between that work and ours.

The SGR model is simple enough to allow a fairly full exploration of
the link between ageing and rheology.  As well as providing some
quantitative predictions of rheological ageing, this allows a broader
discussion of the conceptual framework within which rheological data
for ageing systems should be analysed and interpreted. This conceptual
framework is broader than the SGR model itself; for example it is
known that ageing concepts developed for spin-glass dynamics can also
be applied to problems of domain-growth and
coarsening~\cite{BouCugKurMez98}. Many soft solids, such as polydomain
defect textures in ordered mesophases of copolymers or surfactants,
may show ageing through such coarsening dynamics, or through glassy
rearrangement of domains, or both. While the SGR model is intended to
address only the second feature, the broader conceptual framework we
present can allow for both mechanisms (in which case a superposition
of ageing dynamics with different timescales may result; see
Eq.~\(eqn:general_ageing) below).

Thus we begin in Secs.~\ref{sec:rheology} and~\ref{sec:ageing} by briefly
introducing rheology and
ageing respectively. Then in Sec.~\ref{sec:model} we review the SGR
model, and discuss the
origin of its glass transition and the nature of the glass phase. We also
briefly describe its
rheology under non-ageing conditions; this is discussed fully
elsewhere~\cite{SolLeqHebCat97,long_El}.   In Sec.~\ref{sec:ageing_sgr} we give
a general discussion of
ageing within the SGR model, which sets the stage for our new results for
the linear and nonlinear
rheological response of the SGR model in regimes where ageing cannot be
neglected. The results for
controlled strain conditions are presented and discussed in
Sec.~\ref{sec:results_strain}; those
for controlled stress, in Sec.~\ref{sec:results_stress}. We close in
Sec.~\ref{sec:conclusion}
with a brief summary and our conclusions.

We now discuss the connection between our work and that
of~\citet{Struik78}
on polymeric glasses. Struik presented many
experimental results for such systems, and gave a coherent qualitative
explanation of their ageing in terms of a slow relaxation of the free
volume in the system below the glass point. However, he did not
propose a model constitutive equation for this or any other class of
ageing material. He argued that the effective relaxation time of a
system of age $t_{\rm w}$ (the `waiting time' since sample
preparation) varies $\tau(t_{\rm w}) = t_{\rm w}^\mu \tau_0^{1-\mu}$,
where $\tau_0$ is a microscopic time and $\mu \simeq 1$; apart from
this uniform rescaling of (all) the rheological relaxation time(s),
the material properties are almost invariant in time. (This is his
`time waiting-time superposition' principle; we show below that the
SGR model offers a concrete example of it, with $\mu = 1$.) We do not
expect the SGR model, which makes no mention of the free-volume
concept, to be particularly relevant to polymeric glasses;
nonetheless, various points of contact with Struik's work are
indicated below.
 
\section{Rheology}
\label{sec:rheology} Here we review the basic principles of rheology.
Unlike most in the
literature, our formulation does not assume time translational invariance
(TTI); parts of it may
therefore be unfamiliar, even to rheologists. The formalism allows in
principle an arbitrary
dependence of the material properties on time; we defer to
Sec.~\ref{sec:ageing} a discussion of
what more specific form this dependence might take in materials
which exhibit actual ageing effects (rather than other, more trivial
time dependencies).

\subsection{Constitutive Properties}
\label{sec:con_prop}

Rheology is the study of the deformation and flow properties of materials.
In general, deformation
can comprise volume changes, extensional strain, and shear strain; here we
consider incompressible
materials and assume that only shear strains arise. A system's shear stress
$\sigma(t)$ then depends
functionally on its strain rate history
$\dot\gamma(t'<t)$, where $\dot \gamma$ is the strain rate.  Conversely,
$\gamma(t)$ can be expressed as a functional of the preceding stress
history. A specification of
either type is referred to as a {\em constitutive equation}. In general, of
course, the
constitutive equation is a relationship between stress and strain {\em
tensors}; see~\citet{DoiEdw86} for an introduction. We ignore the tensorial
aspects here, because the
model we describe later is too simple to address them.

\subsection{Step Strain}
\label{sec:step_strain}

A standard rheological test consists of suddenly straining a
previously undeformed material by an amount $\gamma_0$. Suppose this
to be done at time $t_{\rm w}$: then $\gamma(t)=\gamma_0
\Theta(t-t_{\rm w})$, where $\Theta$ is the usual step function. (For
the moment, $t_{\rm w}$ is an arbitrary time label, but later we will
take it as the time that the strain is applied, relative to the
preparation of the sample in some prescribed state, at time zero.) In
general the response can be written
\begin{equation}
\sigma(t) = \gamma_0 G(t-t_{\rm w},t_{\rm w};\gamma_0) \label{eqn:step}
\end{equation}
thereby defining the step strain response,
$G(t-t_{\rm w},t_{\rm w};\gamma_0)$. Note that, by
causality, $G$ vanishes for negative values of its first argument.

\subsection{Time Translation Invariance; Linearity}
\label{sec:lin_tti}

If the material properties of a sample have TTI, then
the time
$t_{\rm w}$ of the initial step strain is irrelevant; the response function
$G(t-t_{\rm w},t_{\rm w};\gamma_0)$
can be written $G(t-t_{\rm w};\gamma_0)$ and depends only
on the elapsed time
since the step strain was imposed.

It is particularly important to recognize that TTI is a quite separate
issue from the {\em
linearity} of the material's response to stress. Even when TTI is absent,
in the small deformation
limit ($\gamma_0 \rightarrow 0$), a regime may exist for which $\sigma$ is
linearly related to
$\gamma_0$:
\begin{equation}
\lim_{\gamma_0\to 0} G(t-t_{\rm w},t_{\rm w};\gamma_0)  = G(t-t_{\rm
w},t_{\rm w})
\label{eqn:steplin}
\end{equation}
The system's stress response is then linearly proportional to strain
amplitude (in the sense that
doubling the strain at all earlier times will cause the stress to be
doubled), even if it also
depends on (say) the sample's age relative to an absolute time of preparation.

Only by assuming {\em both} linearity and TTI do we obtain
\begin{equation}
\sigma(t) = \gamma_0 G(t-t_{\rm w}) \label{eqn:steplintti}
\end{equation}
where the function $G(t)$ is called the time-dependent modulus, or the
linear stress relaxation
function, of the material. If a linear material with TTI is subjected to a
small time-dependent
strain
$\gamma(t)$, then by decomposing this into a sequence of infinitesimal step
strains, one finds
\begin{equation}
\sigma(t) = \int_{-\infty}^t G(t-t')\dot\gamma(t')\, dt' \label{eqn:lincontti}
\end{equation} which is, for a linear material with TTI, the constitutive
equation between stress
and strain. In the steady state ({\em i.e.}, for constant strain rate
$\dot\gamma$) one recovers:
\begin{equation}
\sigma=\dot{\gamma}  \int_0^{\infty} G(t'')  dt'' \label{eqn:viscosity}
\end{equation} The integral, whenever it exists, defines the material's
zero-shear viscosity
$\eta$. For many soft materials, however, $G(t)$ decays to zero so slowly
that the integral
diverges. In this case, there can be no regime of linear response in steady
shear flow, although
there may be a linear regime in, say, oscillatory shear.

Note that there is no unique extension of~\(eqn:lincontti) to the nonlinear
case; only in the
linear regime can one superpose the contributions from each small strain
increment in this fashion.
(In some models, the stress for a general flow is indeed written as an
integral involving the
nonlinear step strain response~\cite{BerKeaZap63};
but this is not generally valid.) On the other
hand,~\(eqn:lincontti) is easily extended to the case where TTI is absent:
\begin{equation}
\sigma(t) = \int_{-\infty}^t G(t-t',t')\dot\gamma(t')\, dt' \label{eqn:lincon}
\end{equation} which represents the most general form of a (nontensorial)
linearized constitutive
equation.

\subsection{Behaviour of the Linear Response Function}
\label{sec:lin_resp_beh}

The principle of causality demands that the response function
$G(t-t_{\rm w},t_{\rm w})$ is zero for times $t<t_{\rm w}$. At
$t=t_{\rm w}$, when strain is applied, $G$ typically increases very
rapidly (in effect discontinuously) to a value $G_0$ which represents
an instantaneous elastic response with modulus $G_0$. Thereafter,
$G(t-t_{\rm w},t_{\rm w})$ is (almost always) a decaying function of
its first argument: the more nearly the material approximates a
viscous liquid, the more rapidly will the stress decay. Specializing
to the TTI case, we recall that for a purely Newtonian liquid of
viscosity $\eta$, the function $G(t)$ approaches a delta function
$\eta\delta (t)$. (This shows that $G_0$ can be infinite so long as
the subsequent decay is rapid enough.) On the other hand an ideally
Hookean elastic material has $G(t)=G_0$, the static shear modulus: in
this case the induced stress will never decay. (Note that properly one
should write $G(t) = G_0\Theta(t)$; the extra factor of $\Theta(t)$,
implied by causality, is omitted here and below.)

Both the Newtonian fluid and the Hookean solid are idealized limiting
cases; most real materials
display behaviour intermediate between these limits and are, on some
timescale at least, {\em
viscoelastic}. For the soft materials of interest to us, the relevant
timescale is readily
observable in rheological experiments. The simplest (TTI) example of
viscoelasticity is the Maxwell
fluid, which is solid-like at short times and liquid at longer ones, with a
simple exponential
response function $G(t) =G_0 \exp(-t/\tau)$ connecting the two. (Its
viscosity obeys
$\eta = G_0\tau$.) This behaviour is seen in a few experimental
systems~\cite{CatCan90}, but  more
often one has $G(t) = G_0 \mu(t)$ where the memory function $\mu(t)$ is not
a single exponential.
In many materials it is possible to identify a longest relaxation time via
$\tau_{\rm max}^{-1} =
-\lim_{t\to\infty} \log
\mu(t)/t$. However, in several important cases, such as a pure power law
relaxation, $\mu(t) \sim
t^{-y}$, the required limit does not exist; the longest relaxation time is
infinite.

\subsection{Creep Compliance}
\label{sec:creep}

Arguing along parallel lines to those developed above, one can in general
write the strain response
to a  step stress $\sigma(t)=\sigma_0\Theta(t-t_{\rm w})$ as
\begin{equation}
\gamma(t)=\sigma_0 J(t-t_{\rm w},t_{\rm w};\sigma_0)
\end{equation}
The linear creep compliance $J(t-t_{\rm w},t_{\rm w})$ is then found by
letting $\sigma_0\to 0$
(assuming this limit exists). This is the main rheological function considered
by~\citet{Struik78} and the one most relevant to studies of ageing in
polymeric glasses (since these are often used, for example, as structural
components subject to steady loads).

In the presence of TTI the linear compliance reduces to a function of
one time variable, $J(t-t_{\rm w})$. For the examples of a viscous
liquid, an elastic solid, and a Maxwell material we have (again
omitting factors of $\Theta(t)$) $J(t)=t/\eta$, $J(t) = 1/G_0$, and
$J(t) =1/G_0+t/\eta$, respectively.  For any material with TTI, the
zero-shear viscosity $\eta$ is defined experimentally as the limiting
ratio of stress to strain rate long after application of an
infinitesimal step stress; it therefore obeys $\eta^{-1} =
\lim_{t\to\infty} dJ(t)/dt$, which may be shown\footnote{The given
limit may also be written $\eta^{-1} = \lim_{\omega \to 0} i\omega
J^*(\omega)$ which, by reciprocity of $J^*$ and $G^*$, implies $\eta =
\lim_{\omega \to 0} G^*(\omega)/i\omega$.  The last definition is
equivalent to~\(eqn:viscosity).  See Sec.~\ref{sec:spectra} for
definitions of $J^*$ and $G^*$.}  to be equivalent
to~\(eqn:viscosity).  A finite viscosity, requires, of course, that
the limit is finite; this is discussed further in
Sec.~\ref{sec:flow_curve} below.

\subsection{Viscoelastic Spectra}
\label{sec:spectra} A common experiment is to apply a steady oscillatory
strain and measure the
resulting stress, or vice versa. For example, the choice

\begin{equation}
\gamma(t)=\Theta(t-t_{\rm s})\mbox{Re}\left[\gamma_0
           e^{i(\phi+\omega t)}\right]
\end{equation} describes an oscillatory flow started at time
$t_{\rm s}$ and continued up to (at least) the time $t$ at which the stress is
measured. Using the linear
constitutive equation for a system with TTI~\(eqn:lincontti), we have
\begin{eqnarray}
\sigma(t)&=&\mbox{Re}\left[\gamma_0 i\omega \int_{t_{\rm s}}^t 
e^{i(\phi+\omega t')} G(t-t')\,dt'
+\gamma_0 e^{i(\phi+\omega t_{\rm s})}G(t-t_{\rm s}) 
\right]\nonumber\\
         &=&\mbox{Re}\left[\gamma_0 e^{i(\phi+\omega t)} \left(
i\omega \int_0^{t-t_{\rm s}} e^{-i\omega t''} G(t'') dt''
+ e^{-i\omega(t-t_{\rm s})}G(t-t_{\rm s}) \right)
\right]   \label{eqn:spectrumts}
\end{eqnarray}
where the second term accounts for any step strain arising at the
switch on time $t_{\rm s}$.
As the number of cycles becomes very large
($\omega(t-t_{\rm s})\gg 1$), transient effects
become negligible, and the stress settles to a simple oscillatory function
of time. In this
steady-state limit we can write
$\sigma(t)=\mbox{Re}\left[G^*(\omega)\gamma(t)\right]$
where\footnote{If $G(t)$ has a non-decaying contribution
$G(t\to\infty)>0$, the form $G^*(\omega)=G(0)+\int_0^\infty
e^{-i\omega t} G'(t)\, dt$, derived
from~(\protect\ref{eqn:spectrumts}) by integration by parts,
should be used instead of~(\protect\ref{eqn:fourierg}). The same relation
can be obtained from~(\protect\ref{eqn:fourierg}) 
by inserting a regularizing factor $e^{-\epsilon t}$
and taking the limit $\epsilon\to 0$.
This corresponds to an oscillatory strain that is switched on by
very slowly increasing its amplitude.}
\begin{equation} G^*(\omega) = i\omega \int_0^\infty e^{-i\omega t} G(t)\, dt
\label{eqn:fourierg}
\end{equation} which is, to within a factor $i\omega$, the Fourier
transform of the stress
relaxation modulus $G(t)$. Traditionally one writes
\begin{equation}
\label{eqn:real_and_im_g} G^*(\omega) = G'(\omega)+iG''(\omega)
\end{equation}
 where $G',G''$ are called respectively the storage and loss moduli of the
material,
and measure the in-phase (elastic) and out-of-phase (dissipative) response
to an
applied strain\footnote{Many commercial rheometers are configured to
deliver the storage and loss
spectra automatically, from a measurement of the amplitude and phase
relations between stress and
strain in steady state.}.

Clearly one can reach an identical steady state by applying a small
amplitude oscillatory stress
and measuring the resulting strain. This defines a new response function
$J^*(\omega)$ via
$\gamma(t) = \mbox{Re} \left[J^*(\omega)\sigma(t)\right]$, which is
evidently just the reciprocal
of
$G^*(\omega)$. But by an argument similar to that given above
for~\(eqn:fourierg) one also has
$J^*(\omega) = i\omega
\int_0^\infty e^{-i\omega t} J(t)\, dt$. Hence, within the linear response
regime of a  system with
TTI, knowledge of any one of $G(t),J(t),G^*(\omega),J^*(\omega)$ is enough
to determine the other
three. (Of course, this ignores any practical limitations on the time and
frequency domains
accessible by experiment.)

Beyond the linear response regime, it is sometimes useful to define
$G^*(\omega;\gamma_0)$ and $J^*(\omega;\sigma_0)$ from the response to a
finite amplitude
oscillatory shear. However, the interest in these quantities is more
limited since, whenever the
strain dependence is nontrivial, there is  no analogue of~\(eqn:fourierg)
relating the nonlinear
oscillatory response to that in step strain or step stress.

\subsection{Viscoelastic Spectra without TTI}
\label{sec:spectra_nontti}

The proper definition of linear viscoelastic spectra for systems without
TTI is more subtle, and is
to some extent a matter of choice. Let us envisage again 
the following idealized
experiment: ($i$) the
sample is prepared in a known state at time zero; ($ii$) a small amplitude
oscillatory shear of
amplitude $\gamma_0$ and phase $\phi$ is started at later time
$t_{\rm s}$, so that $\gamma(t) =
\Theta(t-t_{\rm s})\mbox{Re}\left\{\gamma_0\exp\left[i(\phi+\omega
t)\right]\right\}$;  ($iii$) this is
maintained up to (or beyond) a time
$t$ at which point the stress is measured.  Using the linear constitutive
equation~\(eqn:lincon), we obtain
\begin{eqnarray}
\sigma(t) &=& \mbox{Re}\left[\gamma_0 i\omega \int_{t_{\rm s}}^t 
e^{i(\phi+\omega t')} G(t-t',t')\,dt'
+\gamma_0 e^{i(\phi+\omega t_{\rm s})}G(t-t_{\rm s},t_{\rm s}) 
\right]\nonumber\\
 &\equiv & \mbox{Re}\left[ \gamma_0
e^{i(\phi+\omega t)} G^*(\omega,t,t_{\rm s})
\right]\nonumber
\end{eqnarray} This unambiguously defines a {\em time-varying} viscoelastic
spectrum\footnote{Note that in principle, to identify by experiment
  the real and imaginary parts of $G^*$ for a particular
  $\omega,t,t_{\rm s}$ one would require the experiment to be repeated
  for two different phases $\phi$ (e.g. pure sine and cosine
  deformations). A more common procedure is, of course, to maintain
  the oscillatory strain over many cycles and record the ``steady
  state" amplitude and phase response of the stress. For systems
  without TTI the latter are not uniquely defined.  Only when material
  properties vary slowly enough will this give a definite result;
  whenever it does, it will coincide with~\(eqn:firstgstar).  The
  required conditions are considered, for the SGR model, below.} as
\begin{equation}
G^*(\omega,t,t_{\rm s}) = i\omega \int_{t_{\rm s}}^t 
e^{-i\omega(t-t')} G(t-t',t')\,dt' +
e^{-i\omega (t-t_{\rm s})} G(t-t_{\rm s},t_{\rm s})
\label{eqn:firstgstar}
\end{equation}
A similar compliance spectrum,
$J^*(\omega,t,t_{\rm s})$ can be defined by exchanging stress and
strain in this protocol.

Since it depends on two time arguments as well as frequency,
$G^*(\omega,t,t_{\rm s})$ is a somewhat cumbersome object. However,
simplifications can be hoped for in
the limit
$\omega(t-t_{\rm s}) \gg 1$. In the TTI case, this condition eliminates simple
transients, and allows one
to relate $G^*(\omega)$ to the Fourier transform of $G(t)$
(see~\(eqn:spectrumts)). Corresponding
simplifications are certainly not guaranteed in the absence of TTI.
However, the transient
dependence on $t_{\rm s}$ {\em may} become negligible\footnote{For the
SGR model,
an additional
requirement is that $\omega t_{\rm s}
\gg 1$; see Sec.~\ref{sec:results_strain_linear_osc} below.} when
$\omega(t-t_{\rm s}) \gg 1$, in which case we have
\begin{equation} G^*(\omega,t,t_{\rm s}) \to  G^*(\omega,t)
\label{eqn:secondgstar}
\end{equation} giving a viscoelastic modulus that depends only on the
measurement time $t$. If, in
addition, the time evolution of the underlying material properties is
negligible on the timescale
of one oscillation, then
$G^*(\omega,t)$ {\em may} obey the relation
\begin{equation} G^*(\omega,t) = i\omega \int_0^\infty e^{-i\omega t'}
G(t',t)\, dt'
\label{eqn:maybe}
\end{equation}
Similarly, for $\omega(t-t_{\rm s}) \gg 1$ the compliance spectrum
{\em may} become $t_{\rm s}$-independent,  $J^*(\omega,t,t_{\rm s}) \to
J^*(\omega,t)$, and {\em may} be related to the step stress response via
\begin{equation} J^*(\omega,t) = i\omega \int_0^\infty e^{-i\omega t'}
J(t',t)\, dt'
\label{eqn:maybe2}
\end{equation}
Finally, $G^*(\omega,t)$ and $J^*(\omega,t)$ {\em may} obey the
conventional reciprocal relation $G^*(\omega,t)$ $=$ $1/J^*(\omega,t)$.
Indeed, we shall find that all the above simplifying relationships 
are true for the SGR model studied below. As discussed at the end of
Sec.~\ref{sec:ageing}, they may also hold more generally for
systems with what we term there ``weak long term memory''. However, we do
not have a rigorous proof for this.
Pending such a proof, the above simplifications remain
hypotheses needing
explicit verification for any constitutive model. Experimenters should
likewise beware that, for
systems without TTI, such textbook relationships between the oscillatory
and step strain response
functions cannot be assumed, but should be empirically verified, for each
system studied. 
This {\em prima facie} breakdown of conventional linear viscoelastic
relationships in ageing systems was emphasized by~\citet{Struik78},
though he argued that they are recovered in
sufficiently `short-time' measurements.  It does not (as Struik seems
to suggest) extend necessarily to breakdown of linear superposition
itself, which survives in the form of ~\(eqn:lincon).

\subsection{Steady State Response: The Flow Curve}
\label{sec:flow_curve}

Consider now the ultimate state of a material with TTI long after an
infinitesimal step stress of
amplitude $\sigma_0$ has been applied. The ultimate deformation
may involve a limiting strain $\gamma = \sigma_0 J(t\to \infty)$, in which
case the steady state
(linear) Hookean elastic modulus is $G_\infty=\sigma_0/\gamma$.
Alternatively, the ultimate state
may involve a limiting strain rate, in which case the zero-shear viscosity is
$\eta=\sigma/\dot\gamma$. However, neither outcome need occur. If, for
example, one has ``power law
creep", {\em i.e.}, 
$J(t) \sim t^{y}$ with $0<y<1$, the material has both zero modulus
(infinite compliance) and
infinite viscosity in steady state. There is no rule against this, although
it does require
nonanalyticity of $G^*(\omega)$ at small
frequencies~\cite{SolLeqHebCat97,long_El}, such  that
$\tau_{\rm max}$ is infinite.

What if the stress amplitude is larger than infinitesimal? The ultimate
steady state can again be
that of a solid,  a liquid, or something in between. In cases where a
liquid-like response is
recovered, it is conventional to measure the ``flow curve", which is a
steady state relationship
between stress and strain rate:
\begin{equation}
\sigma_{\rm ss} = \sigma(\dot \gamma)
\label{eqn:flowcurve}
\end{equation}
In many materials, the following limit, called the yield stress
\begin{equation}
\sigma(\dot\gamma \to 0) = \sigma_{\rm y}\label{eqn:yieldstress}
\end{equation}
is nonzero.\footnote{The experimental existence of a true yield stress,
as defined by this limit, is debatable~\cite{BarHutWal89}; behaviour
closely approximating
this is, however, often reported. Note that our definition of yield
stress, from the flow curve, is unrelated to that of~\citet{Struik78}
who defines a `tensile yield stress' at
constant strain rate.}
Note, however, that the presence of nonzero yield stress does not
necessarily imply a
finite Hookean modulus
$G_\infty$: for $\sigma <\sigma_{\rm y}$, the material could creep forever,
but at an ever
decreasing rate.\footnote{Alternatively, it could reach a steady strain
$\gamma$ that is not linear in $\sigma$ even as
$\sigma \to 0$.} Nor does the {\em absence} of a finite yield stress imply
a finite viscosity; a
counterexample is the power law fluid, for which $\sigma \sim
\dot\gamma^p$. This has $\sigma_{\rm
y}=0$ but, for $p<1$, infinite viscosity $\eta =\lim_{\dot\gamma\to
0}\sigma(\dot\gamma)/\dot\gamma$.

We now turn to materials without TTI. For these, no meaningful definition
of ``steady state
response" exists in general. However, in the SGR model considered below,
TTI is restored for
nonzero $\dot\gamma$~\cite{SolLeqHebCat97,long_El}, and this may be
generic for certain types of
ageing~\cite{SolLeqHebCat97,long_El,BouDea95,Kurchan99}.  If so the flow curve,
including the value of the
yield stress $\sigma_{\rm y}$ (but {\em not} the behaviour for $\sigma
<\sigma_{\rm y}$) remains
well-defined as a steady-state property.

\section{Ageing}
\label{sec:ageing}

So far, we have set up a general framework for describing the rheological
properties of systems
without TTI.  Time translation invariance can be broken, in a trivial sense, by
the transients that any system exhibits during equilibration. We now
consider how such
transients can be distinguished from ageing proper. To focus the
discussion, we consider the linear
step strain response function
$G(t-t_{\rm w},t_{\rm w})$. The other response functions introduced
above can be treated similarly.
We define ageing (of the step strain response) as the property
that {\em a significant part of the stress
relaxation takes place on timescales that grow with the age $t_{\rm w}$ of
the system}. If ageing is present, then in order to see the full stress
relaxation we need to
allow the time $t$ at which we observe the stress to be much larger than
the time $t_{\rm w}$ at which
the step strain has been applied. Formally, we need to consider
\begin{equation}
\lim_{t\to\infty} G(t-t_{\rm w},t_{\rm w})
\label{t_gg_delta_t}
\end{equation}
at {\em fixed} $t_{\rm w}$. On the other hand, if there is no ageing, then the
full stress relaxation
is ``visible'' on finite timescales. This means that as long as $\Delta
t=t-t_{\rm w}$ is large enough,
we observe the full stress relaxation whatever the age $t_{\rm w}$
of the system at the time when the
strain was applied. Formally, we can take $t_{\rm w}$ to infinity
first, and then make $\Delta t$
large, which amounts to considering
\begin{equation}
\lim_{\Delta t\to\infty} \lim_{t_{\rm w}\to\infty} G(\Delta t, t_{\rm w}).
\label{t_ll_delta_t}
\end{equation}
In the absence of ageing, the two ways~(\ref{t_gg_delta_t})
and~(\ref{t_ll_delta_t}) of measuring
the final extent of stress relaxation are equivalent, and we have
\begin{equation}
\lim_{t\to\infty} G(t-t_{\rm w},t_{\rm w}) =
\lim_{\Delta t\to\infty} \lim_{t_{\rm w}\to\infty} G(\Delta t, t_{\rm w}).
\label{STM}
\end{equation}
If the system ages, on the other hand, this equality will not hold:
the right-hand side allows only for the decay of stress by
relaxation modes whose
timescale does not diverge with the age of the system, and thus attains a
limit which
includes elastic contributions from all modes that do have age-related
timescales.
It will be different from the left-hand side, which allows for relaxation
processes occurring on all
timescales, and thus attains a limit in which only completely non-decaying
modes contribute.
We therefore adopt the definition
that a systems {\em ages} if at least one of its response functions
violates~(\ref{STM}). By
contrast, we refer to deviations from TTI in other systems (for which all
significant relaxation
processes can essentially be observed on finite timescales) as {\em
  transients}.  We discuss this point further in the context of the
SGR model in Sec.~\ref{sec:results_strain_linear_step}.

Systems that violate~(\ref{STM}) are referred to as having ``long term
memory''~\cite{CugKur95,BouCugKurMez98,CugKur93}. They can be
further subdivided according to the strength of this memory. To illustrate
this distinction,
imagine applying a (small) step strain to a system at time $t_0$ and
switching it off again at some
later time $t_1$. The corresponding stress at time $t>t_1$ is proportional to
$G(t-t_0,t_0)-G(t-t_1,t_1)$. If this decays to zero at large times $t$,
that is, if
\begin{equation}
\lim_{t\to\infty} [G(t-t_0,t_0)-G(t-t_1,t_1)] = 0
\label{WLTM}
\end{equation}
[and~(\ref{STM}) is violated] then we say that the system has ``weak long
term memory'',
otherwise it has ``strong long term memory''.\footnote{There is a
slight subtlety with the definition of long 
term memory for the
linear step stress response. Eq.~\(WLTM), applied literally to
$J(t-t_{\rm w},t_{\rm w})$, suggests that even a Newtonian
fluid with 
$J(t-t_{\rm w},t_{\rm w})\sim t-t_{\rm w}$ has strong long term
memory, because its strain 
``remembers'' stress
applications in the arbitrarily distant past. This is clearly undesirable
as a definition. The problem can
be cured by ``regularizing'' the step stress response: one simply considers
the material in
question ``in parallel'' with an elastic spring with infinitesimal
modulus.}
Although the weakness
condition~(\ref{WLTM}) does not
hold for all response functions in all ageing systems, it seems rather
natural to expect it, in the
rheological context, for most materials of interest. Indeed, a system with
{\em weak} long term
memory eventually forgets any perturbation that was only applied to it
during a finite period.
Thus, the treatment of a sample directly after it has been prepared (by
loading it into the
rheometer, preshearing, etc.)\ will not have a strong impact on the
age-dependence of its
rheological properties. This is the usual experience, and is obviously
needed for the
reproducibility of experiments results; likewise, it means that one can
hope to make theoretical
predictions which are not sensitive to minute details of the sample
preparation. For the SGR model,
any long term memory is indeed weak (as shown in
Sec.~\ref{sec:results_strain_linear_step} below);
we consider this an attractive feature. Note in any case that a rheological
theory for systems with
strong long term memory might look very different from the SGR model.

We have defined ageing as the property that a significant part of the
stress relaxation
$G(t-t_{\rm w},t_{\rm w})$ takes place on timescales that grow
with the age $t_{\rm w}$ of
the system. In the
simplest case, there is only one such growing timescale, proportional to
the age of the
system itself. The (ageing part of the) stress relaxation then becomes a
function of the scaled time
difference $(t-t_{\rm w})/t_{\rm w}$. We will encounter such simple
ageing behaviour in 
the glass phase of the
SGR model, which is discussed below. More complicated ageing scenarios are
possible, however: There
may be several timescales that grow differently with the age of the system.
This can be represented
as
\begin{equation} G(t-t_{\rm w},t_{\rm w}) = \sum_i {\mathcal
G}_i\left[h_i(t)/h_i(t_{\rm w})\right]
\label{eqn:general_ageing}
\end{equation}
where the functions $h_i(t)$ define the different diverging
timescales. If there is only a single term in the sum, with $h(t)=t$,
then the simplest ageing scenario (shown by the SGR model) is
recovered. On the other hand, for $h(t)=\exp(t/\tau_0)$ (where
$\tau_0$ is a microscopic time) one has TTI. The more general form
$h(t)=\exp[(t/\tau_0)^{1-\mu}]$ interpolates between these two
limiting cases (and, for $t-t_{\rm w}\ll t_{\rm w}$, gives Struik's
general `time waiting-time superposition principle'~\cite{Struik78}).
Otherwise, Cugliandolo and Kurchan have shown under fairly mild
assumptions that~(\ref{eqn:general_ageing})
is the most general representation of the 
asymptotic behaviour of step response and
correlation functions in systems with weak long 
term memory~\cite{CugKur94}.

Let us return now to the status of
Eqs.~(\ref{eqn:secondgstar},\ref{eqn:maybe},\ref{eqn:maybe2}). (These concern
the lack of $t_{\rm s}$-dependence in $G^*(\omega,t,t_{\rm s})$, the
Fourier relationship between frequency and
real-time spectra, and the reciprocity between $G^*$ and $J^*$.)
As stated in Sec.~\ref{sec:spectra_nontti} these equations have no
general validity for systems without TTI. Indeed, one can easily
construct theoretical model systems with {\em strong} long term memory
which violate them. On the other hand, we speculate that systems with
{\em weak} long term memory will generically have the
properties~(\ref{eqn:secondgstar},\ref{eqn:maybe},\ref{eqn:maybe2}).
Plausibility arguments can be given to support this
hypothesis~\cite{thesis}, but these do not yet amount to a proof. The
cautionary remarks at the end of Sec.~\ref{sec:spectra_nontti}
therefore still apply.

\section{The SGR model}

\label{sec:model} 

The SGR model is a phenomenological model which captures many of the
observed rheological properties of soft metastable materials, such as
foams, emulsions, slurries and
pastes~\cite{MacMarSmeZha94,KetPruGra88,KhaSchneArm88,%
MasBibWei95,PanRouVuiLuCat96,HofRau93,MasWei95}. It is
based upon Bouchaud's trap model of glassy dynamics, with the addition of
strain degrees of
freedom, and the replacement of  the thermodynamic temperature by an
effective (noise) temperature.
It incorporates only those characteristics deemed common to all soft glassy
materials (SGM's), namely
structural disorder and metastability. We now review its essential features.

We conceptually divide a macroscopic sample of SGM into many mesoscopic
elements. By mesoscopic we
mean large enough such that the continuum variables of strain and stress
still apply for
deformations on the elemental scale, and small enough such that any
macroscopic sample contains
enough elements to allow the computation of meaningful ``averages over
elements''. We then assign
to each element a local strain
$l$, and corresponding stress $kl$, which describe deformation away from
some local position of
unstressed equilibrium relative to neighbouring elements. The macroscopic
stress of the sample as a
whole is defined to be $\lav kl\rav$, where $\lav\,\rav$ denotes averaging
over elements.  Note
that, for simplicity, (shear-) stress and strain are treated as scalar
properties. The model
therefore does not predict, or allow for, the various normal stresses which
can arise in real
materials undergoing nonlinear shear~\cite{DoiEdw86}.

For a newly prepared, undeformed sample, we make the simplest assumption
that $l=0$ for each
element. (Physically, of course, $\lav l\rav=0$ would be sufficient and is
indeed more plausible.)
The subsequent application of a macroscopic strain at rate $\dot\gamma$
causes each element to
strain relative to its local equilibrium state and acquire a non-zero
$l$. For a given element, this continues up to some maximal strain $l_{\rm y}$,
at which point that
element yields, and rearranges into a new configuration of local
equilibrium with local strain $l=0$.\footnote{This ignores possible
``frustration'' effects: an element may not be able to relax to a
fully unstrained equilibrium position due to interactions with
neighbouring elements. Such effects can be incorporated into the
model, but are not expected to affect the results in a qualitative
way~\cite{long_El}.}
Under continued macroscopic straining, the
yielded element now
strains relative to its new equilibrium, until it yields again; its local
strain (and stress)
therefore exhibits a saw-tooth dependence upon time.

The simplest assumption to make for the behaviour between yields is that
$\dot{\gamma}=\dot{l}$: the material deformation is locally
affine~\cite{DoiEdw86}. Yield events
apart, therefore, the SGR model behaves as an elastic solid of spring
constant $k$. Yields confer a
degree of liquidity by providing a mechanism of stress relaxation.

Although above we introduced yielding as a purely strain-induced
phenomenon, we in fact model it as
an ``activated"  process~\cite{SolLeqHebCat97,long_El}. We assume that an
element of yield energy $E =
\frac{1}{2}kl_{\rm y}^2$, strained by an amount $l$, yields with a certain
rate; this defines the probability for yielding in a unit time
interval. We write this rate as $\tau^{-1}$, where
the characteristic yield time 
$\tau=\tau_0\exp\left[(E-\frac{1}{2}kl^2)/x\right]$ is taken to be
the product of an
attempt time and an
activation factor which is thermal in form. This captures the
strain-induced processes described
above since any element strained beyond its yield point will yield
exponentially quickly; but it
also allows even totally unstrained elements to yield by a process of
activation over the energy
barrier $E$. These activation events mimic, within our simplified model,
nonlinear couplings to
other elements (the barrier heights depend on the surroundings, which are
altered by yield events
elsewhere). A more complete model would treat these couplings explicitly.
However, in the SGR model,
which does not,
$x$ is regarded as an effective  ``noise" temperature to model the process.
Because the energy
barriers are (for typical foams, emulsions, etc.) large compared to the
thermal energy $k_BT$, so
are the energy changes caused by these nonlinear couplings, and so to mimic
these, one expects to
need $x$ of order the mean barrier height $\langle E
\rangle$.\footnote{Whether it is fully
consistent to have a noise temperature $x\gg k_BT$ is a debatable feature
of the
model~\protect\cite{SolLeqHebCat97,long_El}; however, we think the results are
sufficiently interesting to
justify careful study of the model despite any uncertainty over its
interpretation. It is also intriguing to note that similar ``macroscopic''
effective temperatures (which remain nonzero even for $k_B T\to 0$)
have recently been found in other theories of out-of-equilibrium
systems with slow dynamics~\cite{Kurchan99,CugKurPel97}.}
Note that the SGR model treats ``noise-induced'' yield events (where
the strain is much below the yield strain $l_{\rm
y}$, {\em i.e.}, where $\frac{1}{2}kl^2\ll E$) and ``strain-induced'' yield
events (where $\frac{1}{2}kl^2\approx E$) in a unified
fashion. We will nevertheless find it useful below to
distinguish between these two classes occasionally.

\begin{figure}[h]
\begin{center}
\epsfig{file=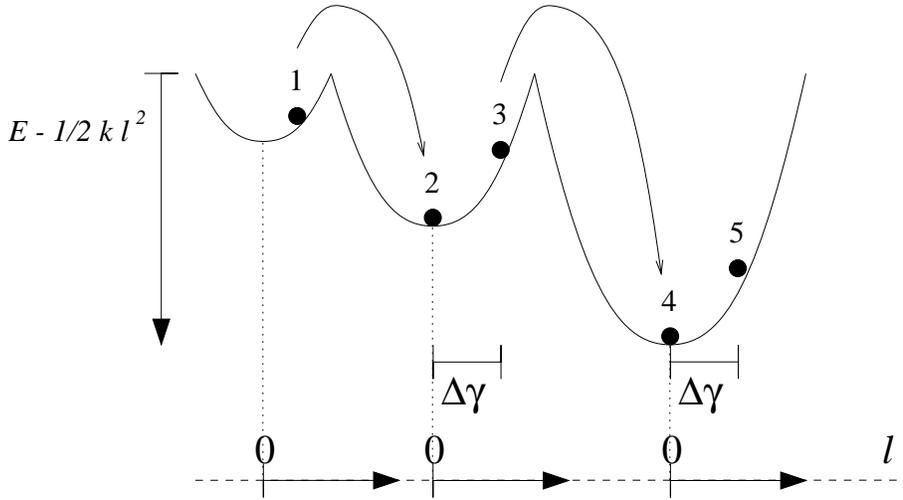,width=12cm}
\end{center}
\caption{Dynamics of the SGR model. A representative particle (1) may hop
out of its trap by
activated hopping ($1\to 2$). It enters the new trap in a state of zero
local strain ($l=0$);
application of strain $\Delta \gamma$ raises its energy ($2\to 3$) making a
subsequent hop ($3\to
4$) more likely. Note that the relative horizontal displacement of the
quadratic potential wells
(traps) is arbitrary; each has its own independent zero for the scale of
the local strain $l$.
\label{fig:wells} } \end{figure}

The disorder inherent to SGM's is captured by assuming that each
element of a macroscopic sample has a different yield energy: a
freshly yielded element is assigned a new yield energy selected at
random from a ``prior'' distribution $\rho(E)$. This suggests the
following alternative view of the dynamics of the SGR model, which is
represented graphically in Fig.~\ref{fig:wells}. Each material element
of a SGM can be likened to a particle moving in a landscape of
quadratic potential wells or ``traps'' of depth $E$. The depths of
different traps are uncorrelated with each other and distributed
according to $\rho(E)$.\footnote{Because of this lack of correlation,
  it does not make sense to think of a particular spatial arrangements
  of the traps.}  The bottom of each trap corresponds to the
unstrained state $l=0$; in straining an element by an amount $l$, we
then effectively drag its representative particle a distance
$\frac{1}{2}kl^2$ up the sides of the trap, and reduce the effective
yield barrier height ($E\rightarrow E-\frac{1}{2}kl^2$). Once the
particle has got sufficiently close to the top of its trap
($E-\frac{1}{2}kl^2 \approx x$), it can hop by activated dynamics to
the bottom of another one. This process corresponds to the yielding of
the associated material element. In the following, we shall use the
terminology of both the ``element picture'' and the ``particle
picture'' as appropriate. Thus, we will refer to
$\tau=\tau_0\exp\left[(E-\frac{1}{2}kl^2)/x\right]$ as either the
yield or relaxation time of an element, or as the lifetime of a
particle in a trap.\footnote{Sometimes this will be further
  abbreviated to ``lifetime of a trap'' or ``lifetime of an
  element''.} The inverse of $\tau$ is the rate at which an element
yields/relaxes or a particle hops. However, we normally reserve the
term yield rate or hopping rate for the {\em average} of these rates
over the whole system, {\em i.e.}, over all elements or particles.
This quantity is denoted $Y$ and will occur frequently below.

A specific choice of $\rho(E)$ is now made:
$\rho(E) =(1/x_{\rm g})\exp(-E/x_g)$, where $x_g = \langle E \rangle$ is
the mean height of a
barrier chosen from the prior distribution $\rho(E)$. As shown
by~\citet{Bouchaud92},  the
exponential distribution, combined with the assumed thermal form for the
activated hopping, is
sufficient\footnote{This is {\em sufficient}, but it is {\em necessary}
only that the
given exponential form be approached at large
$E$.}
to give a glass transition in the model. The transition is at $x =x_g$ and
divides the glass phase
($x\le x_g$), in which weak ergodicity breaking occurs, from a more normal
phase ($x >x_g$). In the
glass phase, the Boltzmann distribution (which is the only possible steady
state for activated
hopping dynamics, in the absence of strain),
\begin{equation} P_{\rm eq}(E) \propto\rho(E)\exp(E/x)
\end{equation}
is not normalizable: thus there is no steady state, and the system must age
with time. (The
converse applies for $x>x_g$: there is then a unique equilibrium state,
which is approached at long
times. Hence ageing does not occur, though there may be transients in the
approach to equilibrium.)
Apart from our use of an effective temperature
$x$, the only modification to Bouchaud's original model of glasses lies in
our introduction of
dynamics within traps coupled to strain.

It may appear suspicious that, to obtain a glass transition at all, an
exponential form of
$\rho(E)$ is required~\cite{Bouchaud92}. In reality, however, the glass
transition is certainly a
collective phenomenon: the remarkable achievement of Bouchaud's model is to
represent this
transition within what is, essentially, a single-particle description. Thus
the chosen ``activated'' form for the
particle hopping rates, and the exponential form 
of the trap depth distribution, should not be
seen as two independent
(and doubtful) physical assumptions, but viewed jointly as a tactic that
allows glassy dynamics to
be modelled in the simplest possible way~\cite{SolLeqHebCat97,long_El}.

From now on, without loss of generality, we choose units so that both $x_g$
and $k$ are unity.
This means that the strain variable $l$ is defined in such a way that an
element, drawn at random from the prior distribution, will yield at strains
of order one.
Since the actual value of the strain variable can be rescaled within the
model (the difference
being absorbed in a shift of $k$), this is purely a matter of convention.
But our choice should be
borne in mind when interpreting our results for nonlinear strains, given
below: where strains ``of
order unity" arise, these are in fact of order some yield strain $l_{\rm y}$,
which the model does not
specify, but which may in reality be a few percent or less. In addition we
choose by convention
$\tau_0 = 1$; the timescale in the SGR model is scaled by the mesoscopic
``attempt time" for the
activated dynamics. The low frequency limit, which is the main regime of
interest, is then defined
by $\omega\tau_0 = \omega \ll 1$. Note that, without our choice of units,
$\langle E\rangle = 1$
so that we expect the interesting physics to involve $x\simeq 1$.

\subsection{Constitutive Equation}

\label{sec:coneq}

The SGR model is exactly solved by two coupled constitutive
equations~\cite{long_El}, the first of
which expresses  strain as an integral over stress history, while the
second embodies the conservation of probability. We assume that the sample
is prepared
(in a known initial state of zero stress and strain) at time zero and that
a time dependent
macroscopic strain
$\gamma(t)$ is applied thereafter, so $\gamma(t)=0$ for $t \le 0$. The
constitutive  equations are
then
\begin{equation}
\label{eqn:ceone}
\sigma(t)= \gamma(t) G_0(Z(t,0)) + \int_0^t
\left[\gamma(t)-\gamma(t')\right]Y(t')G_{\rho}(Z(t,t'))dt'
\end{equation}
\begin{equation}
\label{eqn:cetwo} 1=G_0(Z(t,0)) + \int_0^t Y(t')G_{\rho}(Z(t,t'))dt'
\end{equation} In these equations
\begin{equation}
\label{eqn:Z} Z(t,t')=\int_{t'}^t \exp\left(
          \left[\gamma(t'')-\gamma(t')\right]^2/2x\right)dt''
\end{equation} and $G_\rho(Z)$ and $G_0(Z)$ obey
\begin{equation}
\label{eqn:G_rho} G_{\rho}(Z)=\int_0^{\infty} \rho(E)
\exp\left(-Ze^{-E/x}\right)dE
\end{equation}
\begin{equation}
\label{eqn:G_zero} G_{0}(Z)=\int_0^{\infty} P_0(E)
\exp\left(-Ze^{-E/x}\right)dE
\end{equation} where $P_0(E)$ is the probability distribution for the
yield energies (or trap
depths) in the initial
state of preparation of the sample at time $t=0$. We return below
(Sec.~\ref{sec:sample_preparation}) to the issue of how to choose this
initial state.

These equations can be understood by viewing yielding as a ``birth and
death'' process: each time
an element yields it dies and is reborn with zero stress, and with a yield
energy selected randomly
from the prior distribution $\rho(E)$.  The (average) yield rate
rate at time $t'$ is $Y(t')$; the birth rate at time
$t'$ of elements of yield energy $E$ is therefore $Y(t')\rho(E)$. The
proportion of these which
survive without yielding until time $t$ is found as
$\exp\left[-Z(t,t')/\tau(E)\right]$ where
$\tau(E) = \exp(E/x)$ is the (mean) lifetime that an unstrained
element with yield energy $E$ would have.
The expression~\(eqn:Z) for $Z(t,t')$ reflects the fact that
an element that has last yielded at time $t'$ and has a yield energy
$E$ will have 
a yield rate of $\tau(E)^{-1}
\exp\left(\left[\gamma(t'')-\gamma(t')\right]^2/2x\right)$ at time
$t''$. Here the
exponential factor accounts
for the lowering of the yield barrier by strain applied since the
element last yielded (see
Fig.~\ref{fig:wells}). Note
that this factor is unity under conditions where the local strain is
everywhere negligible, in
which case
$Z(t,t') \to t-t'$ (we return to this point below). More generally,
$Z(t,t')$ can be thought of as an effective time interval measured on an
``internal clock" within an
element, which allows for the effect of local strain on its yield rate, by
speeding up the clock.
This speeding up effect, which describes strain-induced yielding, is the
only source of
nonlinearity within the SGR model.

According to the above arguments, the number  of elements of yield energy
$E$, present at time $t$,
which were last reborn at time
$t'$ is
\begin{equation}
\label{eqn:part_prob_distr_exact} P(E,t,t') =
Y(t')\rho(E)\exp\left[-Z(t,t')/\tau(E)\right]
\end{equation} Such elements each carry a local strain
$\gamma(t)-\gamma(t')$ and so the net contribution they make to the stress
at time
$t$ is
\begin{equation}
\label{eqn:part_stress_distr_exact} s(E,t,t') =
\left[\gamma(t)-\gamma(t')\right]Y(t')\rho(E)\exp\left[-Z(t,t')/\tau(E)\right]
\end{equation}
Integrating these expressions over $t'$ from $0$ to $t$ and adding terms
representing the
contribution from elements which have survived from $t=0$ without yielding
at all, we get
respectively the number
$P(E,t)dE$ of elements at time $t$ with yield energies between $E$ and $E+dE$:
\begin{equation}
\label{eqn:cetwo_E}
\label{eqn:prob_distr_exact} P(E,t)=P_0(E) \exp\left[-Z(t,0)e^{-E/x}\right]
+ \int_0^t P(E,t,t') dt'
\end{equation} and the corresponding stress contribution $s(E,t)dE$ at time
$t$ from such elements:
\begin{equation}
\label{eqn:ceone_E} s(E,t)=\gamma(t)P_0(E)
\exp\left[-Z(t,0)e^{-E/x}\right] + \int_0^t s(E,t,t') dt'
\end{equation}
Integrating~\(eqn:cetwo_E) and~\(eqn:ceone_E) over all yield energies
$E$, we finally recover our constitutive equations~\(eqn:ceone)
and~\(eqn:cetwo) respectively.
Below we will return to these two quantities, which will sometimes be
expressed instead as a
function of the lifetime $\tau(E) = \exp(E/x)$ of an unstrained
element with yield energy $E$, so that $P(\tau,t) d\tau
= P(E,t) dE$ and
likewise for $s$. Note that, because $E\geq 0$, these distributions
are nonzero only for $\tau\geq 1$. We will not write this restriction
explicitly below.

Finally, the following alternative form of the first constitutive
equation~\(eqn:ceone) is sometimes useful:
\be
\sigma(t)=\gamma(t)-\int_0^t\gamma(t')Y(t')G_{\rho}(Z(t,t'))dt'
\label{eqn:modulus_from_ce}
\label{eqn:ceone_alt}
\ee
This is obtained by substituting~\(eqn:cetwo) into~\(eqn:ceone). In
the limit of small strains, $Z(t,t')$ is again replaced by $t-t'$.

\subsection{Rheological Properties of the SGR Model} Solution of the
constitutive
equations~(\ref{eqn:ceone}, \ref{eqn:cetwo})  is relatively straightforward
under conditions where
TTI applies. Here we recall the main results thereby
obtained~\cite{SolLeqHebCat97,long_El}.

\subsubsection{Linear Spectra}
\label{sec:linear_flow_regime} A regime of linear rheological response
arises whenever the effects
of strain on the effective time interval $Z(t,t')$ is small.  This requires
that the local strains
in each element remain small; in  oscillatory shear, where
$\gamma(t) =
\gamma_0 e^{i\omega t}$, this is satisfied at low enough strain amplitudes
$\gamma_0$ for any finite frequency
$\omega$.  (The same is not true in steady shear flow; we return to this in
Sec.~\ref{sec:nonlin}
below.) In the linear regime, the model's internal dynamics are independent
of the imposed
deformation: the elements' lifetimes are, to order
$\gamma_0$, strain-independent. In the constitutive equations, $Z(t,t')$
can then be replaced by
the time interval $t-t'$ (there is no strain-induced yielding).

As described in Sec.~\ref{sec:spectra} above, the conventional definition
of the linear viscoelastic
spectra
$G'(\omega),G''(\omega)$ (Eqs.~\ref{eqn:fourierg},\ref{eqn:real_and_im_g}),
requires not only
linearity but also TTI. Thus they are well-defined only for an equilibrium
state; in the SGR model,
the latter exists only for
$x>1$. But even at $x>1$ these spectra show interesting power law
dependencies at low
frequency\footnote{Here and throughout this paper,
``low frequency" in the
SGR model means,
$\omega \ll 1$, that is, frequencies small compared to the mesoscopic
attempt rate for activated
hopping $\tau_0^{-1} =1$ (in our chosen units).};
these are summarized as follows (the prefactors are
omitted, but discussed
by~\citet{SolLeqHebCat97,long_El}):
\begin{equation}
\label{eqn:spectra}
\begin{array}{lcllcll} G''&\propto &\omega       &\mbox{for $2<x$}, \quad
   &\propto &\omega^{x-1} &\mbox{for $1<x<2$} \\ G' &\propto &\omega ^2
&\mbox{for $3<x$}, \quad
   &\propto &\omega^{x-1} &\mbox{for $1<x<3$}
\end{array}
\end{equation}

Throughout its glass phase ($x\le 1$) where the SGR model violates TTI, we
must study instead the
time dependent spectra $G^*(\omega,t,t_{\rm s})$ as defined  in
Sec.~\ref{sec:spectra_nontti} above; this
is done in Sec.~\ref{sec:results_strain_linear} below. An alternative,
explored
by~\citet{SolLeqHebCat97,long_El,EvaCatSol99} is to observe that TTI can
be restored even for $x
\le 1$ by introducing a cutoff $E_{\rm max}$ in the trap depth distribution
$\rho(E)$. This gives
interesting predictions for $x<1$: for example, one finds  $G'(\omega) \sim
\omega^{1-x}$, for
$\tau^{-1}(E_{\rm max}) \ll
\omega \ll 1$~\cite{SolLeqHebCat97,long_El}.  However, the role of this
cutoff is to bring all
ageing processes to a halt after a large finite time of order
$\tau(E_{\rm max})$; formally there is no long term memory. Since in the
present work we want to study
the ageing regime itself, we assume instead that
$E_{\rm max}$ is infinitely large, so that for $x \le 1$, ageing continues
indefinitely.

\subsubsection{Flow Curve} \label{sec:nonlin}

The flow curve was defined in Sec.~\ref{sec:flow_curve} as the nonlinear
stress response
$\sigma(\gdot)$ to a steady strain rate $\gdot$. For the SGR model, it
shows the
following scalings:

\begin{equation}
\label{eqn:flow_curve}
\begin{array}{lclll}
\sigma   & \propto & \dot{\gamma} &\mbox{for} & x>2      \nonumber   \\
\sigma   & \propto & \dot{\gamma}^{x-1} & \mbox{for} & 1<x<2  \nonumber\\
\sigma-\sigma_{\rm y}  & \propto & \dot{\gamma}^{1-x} & \mbox{for} & x<1 \\
\end{array}
\end{equation}  Here
$\gdot\ll 1$ is assumed; prefactors are discussed by~\citet{long_El}.
The flow curve exhibits
two interesting features which are explored more fully in
Secs.~\ref{sec:steady_shear_nonlin}
and~\ref{sec:results_stress_nonlin_step}. Firstly, for
$x < 1$ there is a yield stress
$\sigma_{\rm y}(x)$ (whose value is plotted in~\cite{long_El}).  A linear
response regime exists at
$\sigma \ll \sigma_{\rm y}$; ageing can occur for all $\sigma <
\sigma_{\rm y}$. For $\sigma > \sigma_{\rm y}$ the system
achieves a steady state, and ageing no longer occurs. This is because any
finite flow rate, however
small, causes strain-induced yielding of elements even in the deepest
traps.\footnote{The time
required to yield, with a steady flow present, is only power law, rather
than exponential in
$E$.} Thus the ageing process is curtailed or ``interrupted" by
flow~\cite{SolLeqHebCat97,long_El};  the flow curve is well-defined (and
independent of the choice
of
$P_0$ in the initial state) even in the glass phase. The second interesting
feature is that, for
$1 < x < 2$ (where ageing is absent) there is no linear response regime at
all in steady shear:
however small the applied stress, the behaviour is dominated by
strain-induced yielding. There is
an anomalous (power law) relation between stress and strain rate, and an
infinite zero-shear
viscosity (cf.\ Sec.~\ref{sec:flow_curve} above). This also shows up
in~\(eqn:spectra), where $\eta
= \lim _{\omega \to 0} G''(\omega)/\omega$ is likewise infinite.

\section{Ageing in the SGR model} 
\label{sec:ageing_sgr}

In this section we discuss some general
features of ageing in the
SGR model; in subsequent ones, we explore the rheological consequences of
these phenomena.

\subsection{Initial Preparation of Sample}
\label{sec:sample_preparation} As noted above, to solve the constitutive
equations~(\ref{eqn:ceone},\ref{eqn:cetwo})  the initial 
distribution
$P_0(E)$ of yield energies or trap depths 
at time zero must be specified. Since we are largely interested in
the rheological
properties of the glass phase ($x \le 1$), for which no steady-state
distribution of yield energies
exists in the absence of flow, we cannot appeal to equilibrium to fix
$P_0(E)$. Instead, this
should depend explicitly on the way the sample was prepared. For
simplicity, we choose the case
where
$P_0(E) = \rho(E)$; this is equivalent to suddenly ``quenching'' the noise
temperature
$x$, at time zero, from a very large value ($x \gg 1$) to a value within
the range of interest. We
refer to it as a ``deep quench".

The question of whether or not a deep quench is a good model for the sample
preparation of a  SGM
remains open~\cite{SolLeqHebCat97,long_El}; since $x$ is not truly a
temperature, it is not clear
exactly how one would realize such a quench experimentally.\footnote{One
argument in its favour is
that it this choice minimizes the information content (maximizes the
entropy) of the initial
distribution
$P_0$; it is therefore a legitimate default choice when no specific
information about the
preparation condition is available.} However, we expect that most
interesting aspects of
ageing behaviour are not too sensitive to the initial quench conditions
$P_0(E)$, so that a deep quench is indeed an adequate model. A study of the
effect of quench depth
on the results for the SGR model is summarized in
App.~\ref{app:finite_temp}; we find independence
of quench depth so long as the final noise parameter $x$ is not too
small.\footnote{More precisely, if the
``deep quench" specification is altered to one in  which, at time zero, the
system is quenched from
equilibrium at $x_0 >1$ to its final noise temperature $x$, 
the leading results are independent
of $x_0$ so long as the
final $x$ value obeys $x>1/(2-1/x_0)$. Note that this condition is
never satisfied for $x<1/2$.}
More generally, a
degree of insensitivity to the initial quench conditions is consistent with
the weak long term
memory scenario; a system whose response decays with a relaxation time of
order its age will
typically lose its memory of the initial state by a power law decay in
time. This can then
easily be swamped by larger, $P_0$ independent contributions, as indeed
occurs in most
regimes of the SGR model (App.~\ref{app:finite_temp}).

Following the initial preparation step, subsequent time evolution of the
rheological response is,
within the glass phase, characterized by an ageing process. To allow
simpler comparisons with the
non-ageing (but still slow) dynamics for $1<x<2$, below we shall also
consider a similar quench
from large $x$ to values lying in this range.

\subsection{Ageing of the Lifetime Distribution}
\label{sec:glass_transition}

We now (following~\citet{Bouchaud92} and~\citet{MonBou96})
discuss in detail the
way ageing affects the lifetime distribution (or equivalently the
distribution of particle hopping rates) within the SGR model.

We ignore the presence of a strain; the following results apply when there
is no flow, and in the
linear response regime, where strain-induced hops can be ignored. Under
such conditions, the
hopping rate $Y(t)$ is a strain-independent function of time, and
is readily found
from~\(eqn:cetwo) by Laplace transform. This is done in
App.~\ref{app:yield_rate}. For the case of
a deep quench (as defined above), the exact asymptotic forms of $Y$ are as
follows:
\begin{equation}
\begin{array}{lclll} Y(t) & = &  \D \frac{x-1}{x}     & \mbox{for}&
x>1\\[5mm] Y(t) & = &  \D
\frac{1}{\ln(t)}  & \mbox{for}& x=1\\[5mm] Y(t) & = &  \D
\frac{t^{x-1}}{x\Gamma(x)\Gamma(1-x)} &
\mbox{for}& x<1\\[5mm]
\end{array}
\label{eqn:hopping_rate}
\end{equation}
where $\Gamma(x)$ is the usual Gamma function. These results assume $t \gg
1$, which we will
usually take to be the case from now on (since timescales of experimental
interest are expected to
be much longer than the mesoscopic attempt time $\tau_0 = 1$). Note that
the late-time asymptotes
given here are subject to various subdominant corrections (see
App.~\ref{app:yield_rate}), some of
which are sensitive to the initial state of sample
preparation.\footnote{For a quench
from initial noise temperature $x_0$, the relative
order of the affected
subdominant terms becomes
$t^{-x(1-1/x_0)}$. Thus, unless one quenches {\em from} a point that is
itself only just above the
glass transition, or {\em to} a point that has $x$ only just above zero,
the exact specification of
the initial state is unimportant at late times.}

A closely related quantity to the hopping rate $Y$ is the distribution of
yield energies
$P(E,t)$ -- which obeys~\(eqn:cetwo_E)
 -- or equivalently the lifetime distribution $P(\tau,t)$. As
previously pointed out, in the absence of strain, the only candidate for a
steady state
distribution of yield energies $P_{\rm eq}(E)$ is the Boltzmann
distribution: $P_{\rm eq}(E)
\propto \rho(E) \exp(E/x)$, which translates to
$P_{\rm eq}(\tau) = P_{\rm eq}(E)dE/d\tau \propto \tau^{-x}$; in either
language, the distribution
is not normalizable for $x<1$, leading to broken TTI in  the
model~\cite{Bouchaud92}.

\begin{figure}[h]
\begin{center}
\epsfig{file=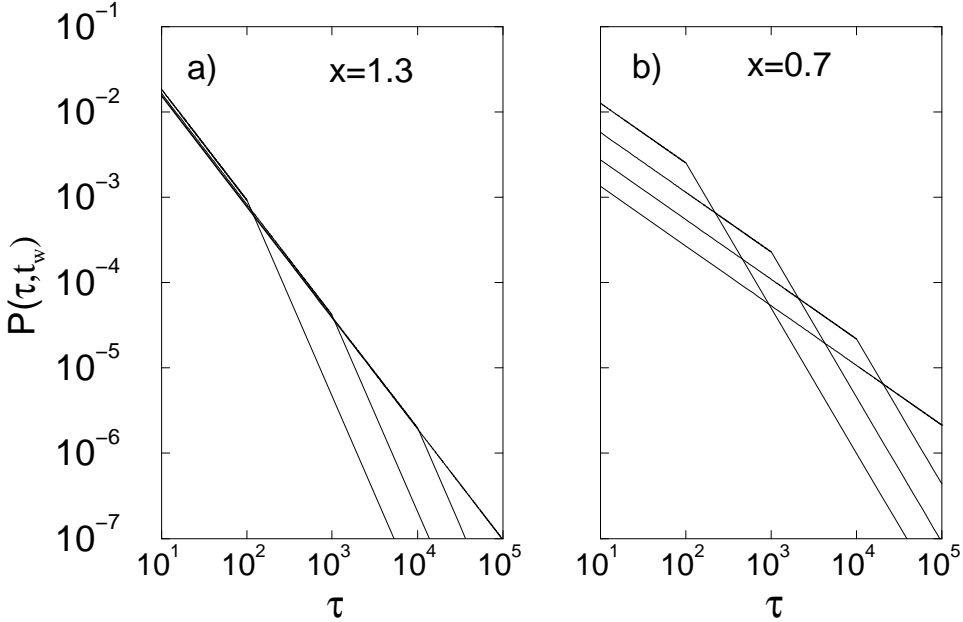,width=12cm}
\end{center}
\caption{Schematic evolution of the relaxation time distribution (a) above the
glass transition; (b)
below it. The first shows a transient decay onto a steady state, the second
shows ageing behaviour.
The curves lie in order of increasing $t_{\rm w}$ at the bottom of each figure.
\label{fig:prob_distr}   }
\end{figure}

Let us therefore consider a deep quench at time $t=0$, and define the
probability distribution for
trap lifetimes $P(\tau,t_{\rm w})$ as a function of the time $t_{\rm
w}$ elapsed since 
sample preparation. (In
Sec.~\ref{sec:results_strain}, we will identify $t_{\rm w}$ with the
onset of a step
strain.) The initial
lifetime distribution, $P(\tau,0)$, describes a state in which the trap
depths are chosen from the
prior distribution $P(E,0) =\rho(E)$; just after a quench to temperature
$x$ the distribution of lifetimes is therefore
$P(\tau,0)\propto\rho(E)d\tau/dE\propto\tau^{-(1+x)}$. Thereafter, by
changing variable from
$E$ to
$\tau$ in~\(eqn:cetwo_E), we find the following approximate expressions for
$P(\tau,t_{\rm w})$
\begin{equation}
\label{eqn:prob_distr_approx}
\begin{array}{lclclcl}
P(\tau,t_{\rm w}) &\simeq & xY(t_{\rm w}) \tau \rho(\tau) &
\mbox{for} &\tau\ll t_{\rm w} &
\mbox{and} & t_{\rm w}\gg 1
\\ P(\tau,t_{\rm w}) &\simeq & xY(t_{\rm w})  t_{\rm w} \rho(\tau) &
\mbox{for} &\tau\gg t_{\rm w}
& \mbox{and} & t_{\rm w}\gg 1
\end{array}
\end{equation}
For a quench temperature above the glass point ($x>1$), $P(\tau,t_{\rm w})$
exhibits a transient decay;
as $t_{\rm w}\rightarrow\infty$, we find (using  the results
in~\(eqn:hopping_rate)) that
$P(\tau,t)\rightarrow P_{\rm eq}(\tau)= (1-x)\tau^{-x}$, as expected. The
nature of the approach to
the long time limit is illustrated schematically in
Fig.~\ref{fig:prob_distr}(a);  the final
distribution has most of its weight at
$\tau=O(1)$, consistent with the  fact that the hopping
rate~\(eqn:hopping_rate) is itself
$O(1)$ in this phase of the model.

For $x<1$, in contrast,
$P(\tau,t_{\rm w})$ evolves as in Fig.~\ref{fig:prob_distr}(b); the limit of
$P(\tau,t_{\rm w})$ is zero for any finite $\tau$ as
$t_{\rm w}\rightarrow\infty$.  Hence, the proportion of elements having yield
time of order unity tends
to zero as
$t_{\rm w}\rightarrow\infty$; the bulk of the  distribution's weight
is at $\tau \simeq
t_{\rm w}$.\footnote{More formally, for
$x<1$, we have $
\lim_{t_{\rm w}\rightarrow\infty}\int_{1}^{b}P(\tau,t_{\rm w}) d\tau
=0$ for any $b>1$, while for any
$a<1<b$ we have instead
$ \lim_{t_{\rm w}\rightarrow\infty}\int_{at_{\rm w}}^{bt_{\rm w}}
P(\tau,t_{\rm w}) d\tau=O(1)$.}
This is consistent
with the decay of the
hopping rate as a power law of $t_{\rm w}$, and with the idea that, in a system
undergoing ageing, the
characteristic relaxation time is typically
of the order the age of the system itself.

\subsection{Higher Moments}
\label{sec:higher_transition} The above analysis focuses on the
time-evolution of the distribution
of elements' lifetimes, which is the usual quantity of interest in the
formal analysis of ageing
effects~\cite{BouDea95,BouCugKurMez98}. Indeed, the latter are usually
attributed to the divergence
of the normalization integral, or zeroth moment, of
$P_{\rm eq}(\tau)$ (undefined, within the Boltzmann distribution, when
$x\le 1$). Formally,
however, one can  consider a series of critical $x$ values, $x_n=n+1$,
below each of which the
$n$th moment of
$P_{\rm eq}$ becomes undefined~\cite{EvaCatSol99,odagaki95}. For $n>0$ this
does not lead to
ageing, in the sense defined in Sec.~\ref{sec:ageing} above, but can lead
to anomalous, slow time
evolution in any experimental measurement that probes the $n$th moment. For
example, in
Sec.~\ref{sec:shear_start_up_linear} below, we discuss the time-evolution
of the distribution of
stresses borne by elements in a steady-shear startup experiment. In steady
state, the stress carried
by an element whose lifetime is $\tau$ is of order $\gdot\tau$.  If
$P(\tau) = P_{\rm eq}(\tau)$ and is unperturbed by flow (as a linear
response analysis would
assume), then the zero-shear viscosity is of order
$\int\tau P_{\rm eq}(\tau) d\tau$, which diverges for $x<2$ (see
Sec.~\ref{sec:flow_curve} above).

\section{Rheological Ageing: Imposed Strain}
\label{sec:results_strain}

In this and the next sections, we describe our new rheological results for
the SGR model. We focus
particularly on rheological {\em ageing}, which occurs in the glass phase
($x<1$); however, several new
results for $1<x<2$, including anomalous {\em transient behaviour}, are also
presented. The case $x=1$,
which divides these regimes, shows its own especially weak (logarithmic)
form of ageing and is,
where necessary, treated separately below.

For simplicity, we consider (for all $x$ values) only the idealized route
to sample preparation
described in Sec.~\ref{sec:sample_preparation} above: the system is
prepared at time
$t=0$ by means of a deep quench, so that $G_0(Z(t,0))$ = $G_\rho(Z(t,0))$
in the constitutive
equations~(\ref{eqn:ceone}, \ref{eqn:cetwo}).  Note that these
constitutive equations for the SGR
model are more readily solved to find the stress response to an imposed
strain, rather than
vice-versa. Accordingly, we focus first on strain-controlled experiments
and defer to
Sec.~\ref{sec:results_stress} our analysis of the stress-controlled case.

\subsection{Linear Response}
\label{sec:results_strain_linear}

As described in Sec.~\ref{sec:linear_flow_regime} above, when local strains
are negligible,
the SGR
model displays a linear response regime. The effective time  interval
$Z(t,t')$ in
Eqs.~(\ref{eqn:ceone},\ref{eqn:cetwo})  becomes the  actual time interval
$t-t'$, and the hopping
rate $Y(t')$ a strain-independent function of time. For the deep quench
considered here, $Y(t')$
assumes the asymptotic 
forms summarized in~\(eqn:hopping_rate). The stress response to
any strain history then
follows  simply from~\(eqn:ceone), by integration.

\subsubsection{Step Strain}

\label{sec:results_strain_linear_step}

For a step strain, the amplitude $\gamma_0$ gives the maximum local strain
experienced by any
element. The condition for linearity in this case is therefore simply
$\gamma_0 \ll 1$.  The
linearized step strain response was defined 
in~\(eqn:steplin).  It is
found for the SGR model\footnote{Note that by construction of the SGR
model, the linear step strain response is actually identical to the
correlation function defined by Bouchaud for his trap
model~\protect\cite{Bouchaud92,MonBou96}.} using~\(eqn:ceone_alt):
\begin{equation}
\label{eqn:lin_step} G(t-t_{\rm w},t_{\rm w})=1-\int_{t_{\rm w}}^t
Y(t')G_{\rho}(t-t')dt'
\end{equation}
 As outlined in App.~\ref{app:step_strain}, analytic limiting forms
for $G(t-t_{\rm w},t_{\rm w})$ can be found when experimental
timescales are large on the scale of the mesoscopic attempt $\tau_0
=1$, so that $t-t_{\rm w}\gg 1$ and $t_{\rm w}\gg 1$. In this limit we
identify two distinct regimes: a short time interval regime $t-t_{\rm
w}\ll t_{\rm w}$ and long time interval regime $t-t_{\rm w}\gg t_{\rm
w}$ (where the measure of ``short" and ``long" is not now $\tau_0$ but
$t_{\rm w}$ itself). The limiting forms in each case depend on the
value of $x$; our results are summarized in table~\ref{table:modulus}.

\renewcommand{\arraystretch}{2.7}

\begin{table}[htb!]
\begin{center}
\begin{tabular}{|l||c|c|}\hline
    & $\D G(t-t_{\rm w},t_{\rm w})\ \ \mbox{for}\ \ t-t_{\rm w}\ll t_{\rm w}$
    & $\D G(t-t_{\rm w},t_{\rm w})\ \ \mbox{for}\ \ t-t_{\rm w}\gg t_{\rm w}$
          \\[5mm]\hline\hline
$x>1$ & $\D \Gamma (x)\left(t-t_{\rm w}\right)^{1-x} $
        & $\D (x-1)\Gamma(x)\frac{t_{\rm w}}{\left(t-t_{\rm w}\right)^x}$
          \\[5mm] \hline
$x=1$   & $\D 1- \frac{\ln\left(t-t_{\rm w}\right)}
             {\ln\left( t_{\rm w}\right)}$
        & $\D \frac{t_{\rm w}}{t-t_{\rm w}}
         \frac{1}{\ln\left(t_{\rm w}\right)}$
          \\[5mm] \hline
$x<1$   & $\D 1- \frac{1}{\Gamma(2-x)\Gamma(x)}
           \left(\frac{t-t_{\rm w}}{t_{\rm w}}\right)^{1-x}$
    & $\D \frac{1}{\Gamma(1+x)\Gamma(1-x)} \left(\frac{t_{\rm
w}}{t-t_{\rm w}}\right)^x$ 
          \\[5mm] \hline
\end{tabular}
\caption{Stress response to step strain  at short and long times ($t-t_{\rm w}
\gg 1$, $t\gg 1$ assumed).
$\Gamma(x)$ denotes the usual Gamma function.
\label{table:modulus}
}
\end{center}
\end{table}

\renewcommand{\arraystretch}{1}

The asymptotic scalings apparent in the various entries of
table~\ref{table:modulus} can be physically motivated by the following
simple arguments. Upon the application of the strain at time $t_{\rm
w}$ the local strain of each element exactly follows the macroscopic
one, and the instantaneous response is elastic\footnote{This is a
general characteristic of the SGR model: whenever the macroscopic
strain changes discontinuously by an amount $\Delta\gamma$, the stress
$\sigma$ also increases by $\Delta\gamma$.}: $G(0,t_{\rm w})=1$. In
the time following $t_{\rm w}$, elements progressively yield and reset
their local stresses $l$ back to zero. The stress remaining at $t$
will be that fraction of elements which has survived from $t_{\rm w}$
without yielding, and hence roughly that fraction $\int_{t-t_{\rm
w}}^{\infty}P(\tau,t_{\rm w}) \, d\tau$ which, at time $t_{\rm w}$,
had time constants greater than $t-t_{\rm w}$.  Hence in measuring the
linear response to a step strain we are probing the properties of the
system as they were at the time of strain application.

\begin{figure}[t]
\begin{center}
\epsfig{file=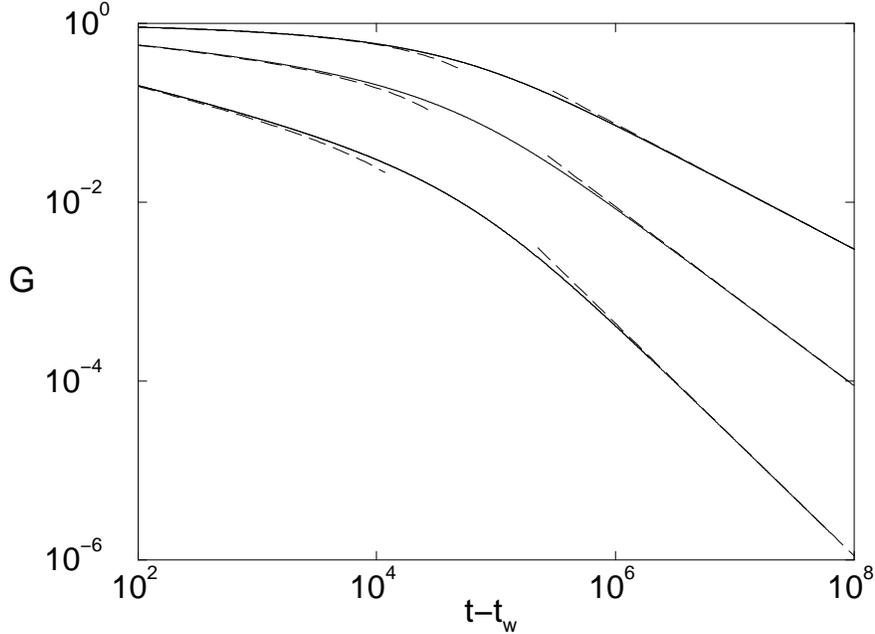,width=11.5cm}
\end{center}
\caption{   
  Approximate curves for $G(t-t_{\rm w},t_{\rm w})$ generated using the
  interpolating
  formulae~(\ref{eqn:G_step_better_fit_one},\ref{eqn:G_step_better_fit_two})
  (dashed lines),
  compared to numerical data for this
  quantity (solid lines). 
  Solid curves downwards show numerical data for $G$ vs
  $t-t_{\rm w}$ at noise temperatures $x=0.7$, $x=1.0$ and $x=1.3$
  respectively. The waiting time $t_{\rm w}$ is $10^5$.
\label{fig:Gstep_formula}}
\end{figure}

Using the approximate expressions given in~\(eqn:prob_distr_approx) above,
we have
$P(\tau,t_{\rm w})\propto
\tau^{-x}$ for
$\tau \ll t_{\rm w}$ and $P(\tau,t_{\rm w})\propto t_{\rm
w}\tau^{-(1+x)}$ for $\tau \gg 
t_{\rm w}$. This gives, for short
time intervals ($t-t_{\rm w} \ll t_{\rm w}$)
\[
G(t-t_{\rm w},t_{\rm w})\simeq1 - \int_1^{t-t_{\rm w}}
P(\tau,t_{\rm w})\, d\tau
      \simeq 1 - x\,
\frac{(t-t_{\rm w})^{1-x}-1}{t_{\rm w}^{1-x}-x}
\]
and, for long time intervals ($t-t_{\rm w} \gg t_{\rm w}$)
\[
G(t-t_{\rm w},t_{\rm w}) \simeq  \int_{t-t_{\rm w}}^{\infty}
P(\tau,t_{\rm w})\, d\tau
         \simeq \frac{(1-x)t_{\rm w}(t-t_{\rm w})^{-x}}{t_{\rm w}^{1-x}-x}
\]
In fact these estimates approximate the numerical data already quite
well. Even better agreement is obtained by adjusting the prefactors to
fit the asymptotic results in table~\ref{table:modulus}:
\begin{eqnarray}
G(t-t_{\rm w},t_{\rm w}) & \simeq &  
1 - \frac{\Gamma(x)(t-t_{\rm w})^{1-x}-1}
{\Gamma^2(x)\Gamma(2-x)t_{\rm w}^{1-x}-1} \qquad \mbox{for } 
t-t_{\rm w} \ll t_{\rm w}
\label{eqn:G_step_better_fit_one}
\\
 G(t-t_{\rm w},t_{\rm w}) & \simeq &
\frac{(x-1)\Gamma(x)t_{\rm w}(t-t_{\rm w})^{-x}}
{1-\Gamma(x)\Gamma(x+1)\Gamma(2-x)t_{\rm w}^{1-x}} \qquad \mbox{for } 
t-t_{\rm w} \gg t_{\rm w}
\label{eqn:G_step_better_fit_two}
\end{eqnarray}
In the relevant time regimes, these formulae agree well with our
numerical results (see Fig.~\ref{fig:Gstep_formula}), at least over
the noise temperature range 0 to 2; they could therefore be used in a
standard curve fitter for comparison with experimental data. In the
limit $t-t_{\rm w}\rightarrow\infty$ and $t_{\rm w}\rightarrow\infty$,
they reproduce (by construction) the results shown in
table~\ref{table:modulus} for $x>1$ and $x<1$. The logarithmic terms
at the glass point $(x=1)$ can also be recovered by taking the limit
$x\rightarrow 1$ first.

\begin{figure}[p]
\begin{center}
\epsfig{file=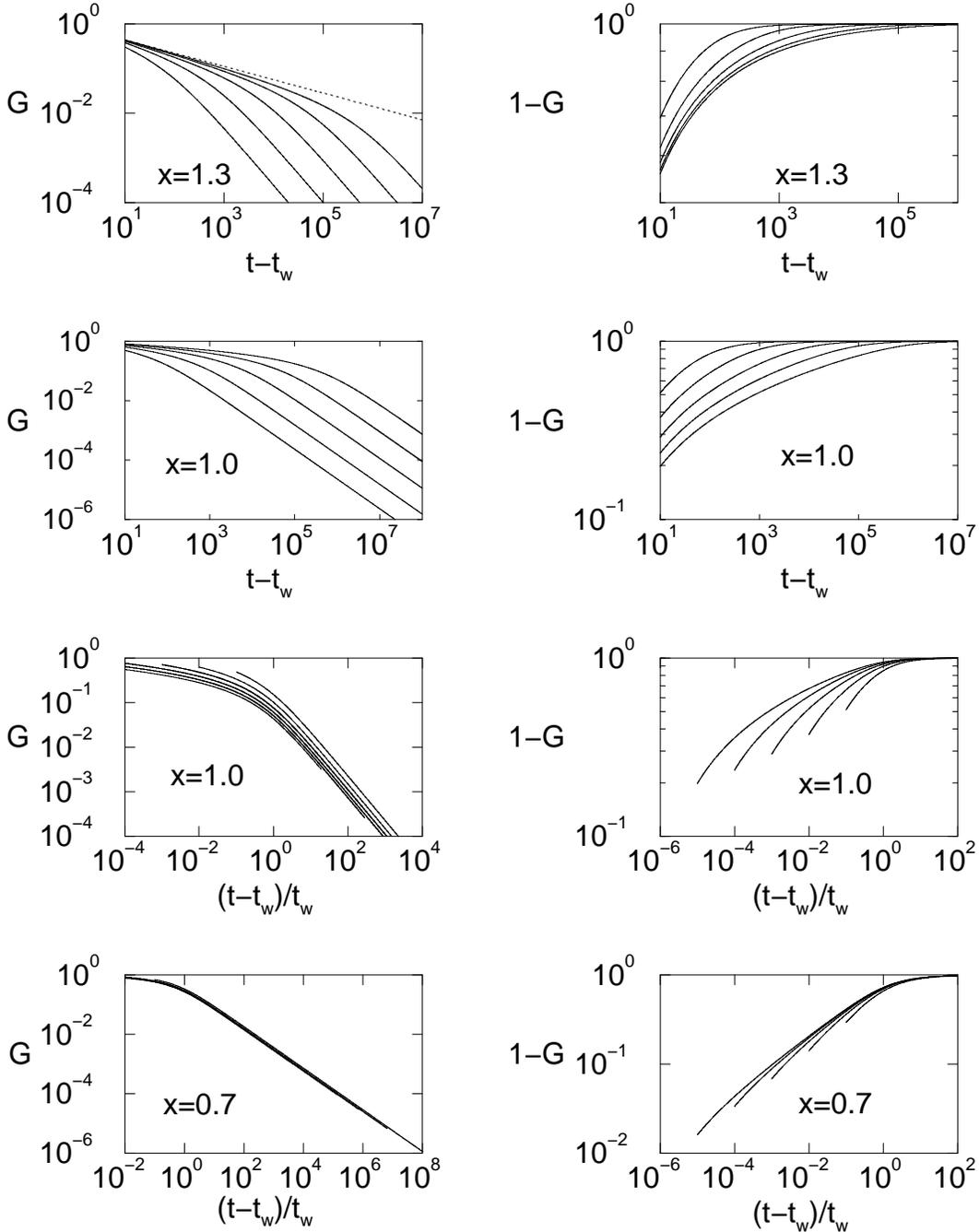,width=14cm}
\end{center}
\caption{Left column: age-dependent stress relaxation modulus
  $G(t-t_{\rm w}, t_{\rm w})$ against scaled time interval $(t-t_{\rm
    w})/t_{\rm w}$ (for $x\le1 $) and time interval $t-t_{\rm w}$ (for
  $x\ge 1$). Right column: $1-G(t-t_{\rm w}, t_{\rm w})$, plotted
  similarly.  Shown are data for waiting times $t_{\rm w}=10^2\,, 10^3
  \ldots 10^6$ (left to right for top four graphs, right to left for
  bottom four graphs). Transients are visible in the top left figure,
  as follows: The curves coincide at short time intervals $t-t_{\rm w}
  \ll t_{\rm w}$.
  At large $t_{\rm w}$, this regime accounts for more
  and more of the decay of $G$; the remaining $t_{\rm w}$-dependence
  is only through an unimportant tail. For $t_{\rm w}\to\infty$, the
  ``short time'' regime extends to all finite values of $t-t_{\rm w}$; one
  recovers the equilibrium response (shown as the dotted line) which
  decays to zero on a $t_{\rm w}$-independent timescale. Ageing is visible
  at bottom left, where the major part of the decay of $G$ occurs on
  a timescale of order $t_{\rm w}$ itself, with unimportant
  corrections to this scaling at early times.
\label{fig:G_step} }
\end{figure}

Using the forms for $G(t-t_{\rm w},t_{\rm w})$ as summarized in
table~\ref{table:modulus}, and substituting these in
Eqs.~(\ref{STM},\ref{WLTM}), we see that the SGR model has short term
memory for $x>1$ and weak long term memory for $x \le 1$. Thus we
expect transients for $x>1$ and ageing for $x\le 1$. As elaborated in
Fig.~\ref{fig:G_step}, this is indeed what we find.  More generally,
these step strain results for the SGR model show some interesting
features of rheological ageing.  Consider first the behaviour above
the glass transition ($x>1$). Here the stress decay at short time
intervals ($t-t_{\rm w}\ll t_{\rm w}$) depends only upon the time
interval between stress imposition and measurement itself ($t-t_{\rm
w}$), and not on the sample age $t_{\rm w}$. This is because the traps
which contribute to stress decay during this interval are mainly those
with lifetimes $\tau < t-t_{\rm w}$; and the population of these traps
has already reached Boltzmann equilibrium before the strain is
switched on (see Fig.~\ref{fig:prob_distr}(a)). Taking the limit
$t_{\rm w}\rightarrow\infty$ at constant $t-t_{\rm w}$, ({\em i.e.},
letting the system fully equilibrate {\em before} we apply the
strain), we recover a TTI stress relaxation function which decays to
zero on timescales of order one (the mesoscopic attempt time).  On the
other hand, for any finite waiting time $t_{\rm w}$, the stress decay
at long enough times ($t-t_{\rm w}\gg t_{\rm w}$) violates TTI, since
it is controlled by decay out of deep traps ($\tau \gg t_{\rm w}$)
which had not already equilibrated before $t_{\rm w}$. Note that even
though this feature of the stress relaxation depends explicitly on
$t_{\rm w}$, it is not an ageing effect according to our definition in
Sec.~\ref{sec:ageing}. This is because the deviations from TTI and the
dependence on $t_{\rm w}$ manifest themselves at ever smaller values
of $G$ as $t_{\rm w}$ becomes large. Equivalently, if we assume that
$G(t-t_{\rm w},t_{\rm w})$ can be measured reliably only as long as it
remains greater than some specified value (a small fraction $\epsilon$
of its initial value $G(0,t_{\rm w})=1$, for example), then the
results will become $t_{\rm w}$-independent for sufficiently large
$t_{\rm w}$.

Below the glass point ($x \le 1$) we see true ageing, rather than
anomalous transient effects: A significant part of the stress
relaxation $G(t-t_{\rm w},t_{\rm w})$ now takes place on timescales
that increase with the sample age $t_{\rm w}$ itself. In fact, in the
case of the SGR model, this applies to the {\em complete} stress
relaxation, and $t_{\rm w}$ itself sets the relevant timescale: for
$x<1$, $G$ depends on time only through the ratio $(t-t_{\rm
w})/t_{\rm w}$.\footnote{This is typical, but not automatic for ageing
systems; the case $x=1$, for example, does not have it.  In general,
the timescale for ageing can be any monotonically increasing and
unbounded function of $t_{\rm w}$. There can also be {\em parts} of
the stress relaxation which still obey TTI. An example is $G(t-t_{\rm
w},t_{\rm w})=g_1(t-t_{\rm w}) + g_2((t-t_{\rm w})/t_{\rm w})$, which
exhibits ageing when $g_2$ is nonzero, but also has a TTI short time
part described by $g_1$.  Superpositions of relaxations with different
ageing timescales are also possible; compare
Eq.~(\protect\ref{eqn:general_ageing}).}  It is still true that stress
decay during the interval $t-t_{\rm w}$ is dominated by traps for
which $\tau <t-t_{\rm w}$, but no longer true that these traps have
reached Boltzmann equilibrium by time $t_{\rm w}$: in an ageing system
such equilibrium is never attained, even for a subset of shallow traps
(see Fig.~\ref{fig:prob_distr}(b)).  Instead, the population of such
traps will gradually deplete with age, as the system explores
ever-deeper features in the energy landscape.  Decay from these deep
traps becomes ever slower; the limit $t_{\rm w}\rightarrow\infty$ (for
any finite $t-t_{\rm w}$) gives completely arrested behaviour in which
all dynamics has ceased, and the system approaches a state of perfect
elasticity ($G=1$). Even in an experiment that can only resolve values
of $G$ above a threshold $\epsilon$ (see above), we would detect that
the stress relaxation becomes slower and slower as $t_{\rm w}$
increases.

The fact that $G$ depends on time only through the ratio
$(t-t_{\rm w})/t_{\rm w}$ is a simple example of Struik's `time
ageing-time superposition' principle~\cite{Struik78}: the
relaxation curves for different $t_{\rm w}$ can be superposed by a
rescaling of the time interval $t-t_{\rm w}$ by the sample
age. However, as mentioned previously, Struik's discussion allows a
more general form in which the scale factor varies as $t_{\rm
w}^\mu$, with $\mu <1$. The case $\mu = 1$, exemplified by the SGR
model, is the only one in which, even at very long times, the
relaxation time does not become short compared to system age.

\subsubsection{Oscillatory Strain}
\label{sec:results_strain_linear_osc}

\begin{figure}[t]
\begin{center}
\epsfig{file=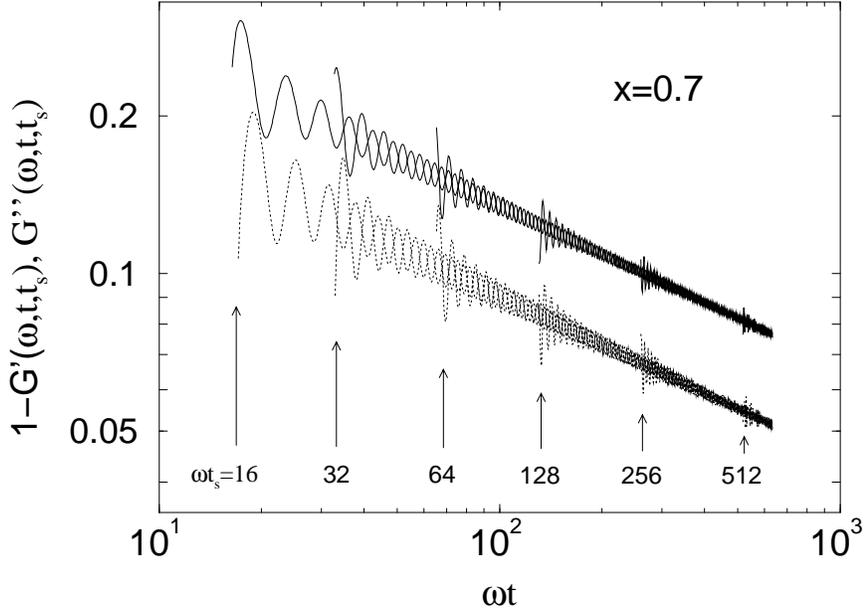,width=11.5cm}
\end{center}
\caption{Demonstration of $t_{\rm s}$-independence
$G^*(\omega,t,t_{\rm s})$ in the 
glass phase. Shown are
$1-G'(\omega,t,t_{\rm s})$ (solid lines) and $G''(\omega,t,t_{\rm s})$
(dotted lines) 
against $\omega t$. The
noise temperature is $x=0.7$ and the frequency
$\omega=0.01$. Start-time values $t_{\rm s}$ obey
$\omega t_{\rm s}=2^4\,, 2^5\,, \ldots 2^9$. When $\omega (t-t_{\rm s})\gg 1$
(a sufficient number of oscillations after the beginning of each
dataset) and $\omega t_{\rm s}\gg 1$ (datasets beginning further on the
right), $G^*$ becomes
independent of $t_{\rm s}$.
\label{fig:Gosc_tsdep} }
\end{figure}

In an oscillatory strain, the maximal local strain of any element is
$\gamma_0$, the strain amplitude. Thus a linear regime in the SGR
model is ensured whenever $\gamma_0 \ll1$. The linear viscoelastic
spectrum, as defined in~\(eqn:firstgstar), can be found for the SGR
model using~\(eqn:modulus_from_ce):
\begin{equation}
\label{eqn:G_osc_lin}
G^*(\omega,t,t_{\rm s})=1-\int_{t_{\rm s}}^{t}e^{-i\omega(t-t')}
Y(t')G_{\rho }(t-t')dt'
\end{equation}
In principle, this quantity depends on $t_{\rm s}$, the time when the
oscillatory strain was started.
However, when the experimental timescales become large, we
find (as shown in
App.~\ref{app:ts_indep}) that this dependence on $t_{\rm s}$ 
is weak. In
fact, within the SGR model, the conditions needed to make $G^*$
negligibly dependent on $t_{\rm s}$ (for low
frequencies, $\omega \ll 1$) are that
$\omega(t-t_{\rm s}) \gg 1$ and $\omega t_{\rm s} \gg 1$. The first
signifies merely 
that many cycles of
oscillatory strain are performed before the stress is measured; the second
ensures that transient
contributions from the initial sample preparation stage (the quench at
$t=0$) are negligible.
Notably, these criteria do not depend on the noise temperature $x$, and
therefore hold even in the
glass phase ($x\le 1$); see Fig.~\ref{fig:Gosc_tsdep}. The fact that they
are sufficient even in
the glass phase is far from obvious physically, and requires a careful
discussion: we give this  in
App.~\ref{app:ts_indep}. Broadly speaking, these criteria are satisfied in
any experiment that would
reliably measure a conventional $G^*(\omega)$ spectrum for systems with TTI.

\begin{figure}[t]
\begin{center}
\epsfig{file=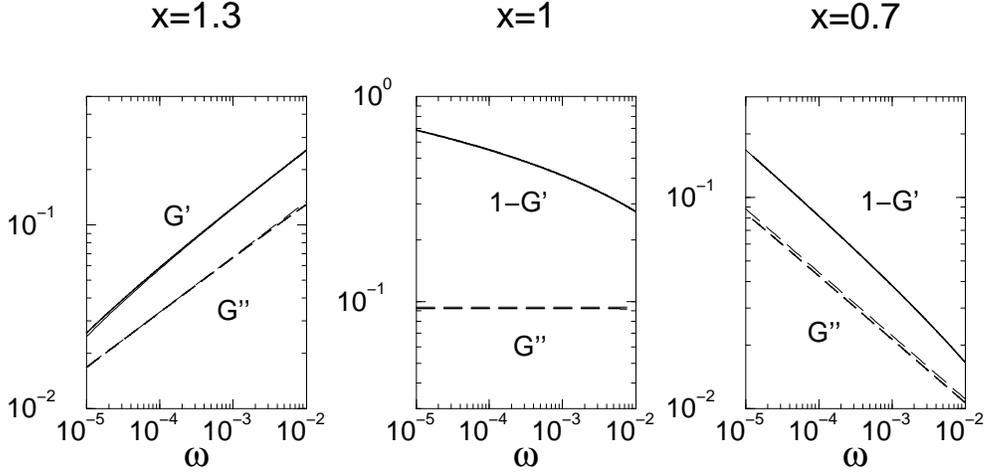,width=13cm}
\end{center}
\caption{Approximate curves for $G'(\omega,t)$ and $G''(\omega,t)$
 generated using the simple interpolating formula~\(eqn:Gosc_formula)
 (thin lines) compared to the numerical data for these quantities
 (thick lines), plotted as a function of $\omega$ at fixed $t=10^7$.
 Noise temperatures $x$ are as shown.  Note that the predictions of the
 interpolating formula are practically indistinguishable from the
 numerical data over the frequency window shown.
\label{fig:Gosc_formula}}
\end{figure}

For the purposes of such experiments, we can therefore drop the $t_{\rm s}$
argument and define a
time-dependent spectrum
$G^*(\omega,t)$. Our results for the long-time behaviour ($t\gg 1$)
of this quantity are as follows (see App.~\ref{app:osc_strain}):
\be
\begin{array}{lclll} G^*(\omega,t) & = & \D \Gamma(x)\Gamma(2-x)
(i\omega)^{x-1}
                      &\mbox{for} & 1<x<2 \\[5mm] G^*(\omega,t) & = & \D
1+\frac{\ln(i\omega)}{\ln(t)}
                      &\mbox{for} & x=1 \\[5mm] G^*(\omega,t) & = & \D
1-\frac{1}{\Gamma(x)}(i\omega t)^{x-1}
                      &\mbox{for} & x<1 \\[5mm]
\end{array}
\label{eqn:G_osc_lin_results}
\ee
For comparison with experimental results, the simple interpolating form
\be
\label{eqn:Gosc_formula}
G^*(\omega,t)=1-\frac{\Gamma(x)\Gamma(2-x)\left(i\omega\right)^{x-1}-1}
{\Gamma^2(x)\Gamma(2-x)t^{1-x}-1}
\ee
may be useful; we have checked that it provides 
a good fit to our numerical data, at least over the
noise temperature range 0 to approximately 1.3 (see
Fig.~\ref{fig:Gosc_formula}).

By measuring $G^*(\omega,t)$ we are directly probing the properties of
the system at the time of measurement, $t$. In light of this, the
results of~\(eqn:G_osc_lin_results) are easily understood.  In the
ergodic phase ($x>1$), $G^*(\omega,t)$ will reach a $t$-independent
value within a time of $O(1/\omega)$ after the quench, as the relevant
traps will then have attained their equilibrium population. The
relaxation time is then of $O(\tau_0)$ (that is, $O(1)$ in our units)
and the response $G^*(\omega,t)$ is a function only of $\omega$. In
contrast, below the glass point the characteristic relaxation time at
the epoch of measurement is of order $t$, and the response is a
function only of the product $\omega t$. Since the losses in an
oscillatory measurement arise from traps with lifetimes less than
about $1/\omega$ (elements in deeper traps respond elastically), the
overall response becomes more elastic as the system ages into traps
with $\tau>1/\omega$.

\begin{figure}[p]
\begin{center}
\epsfig{file=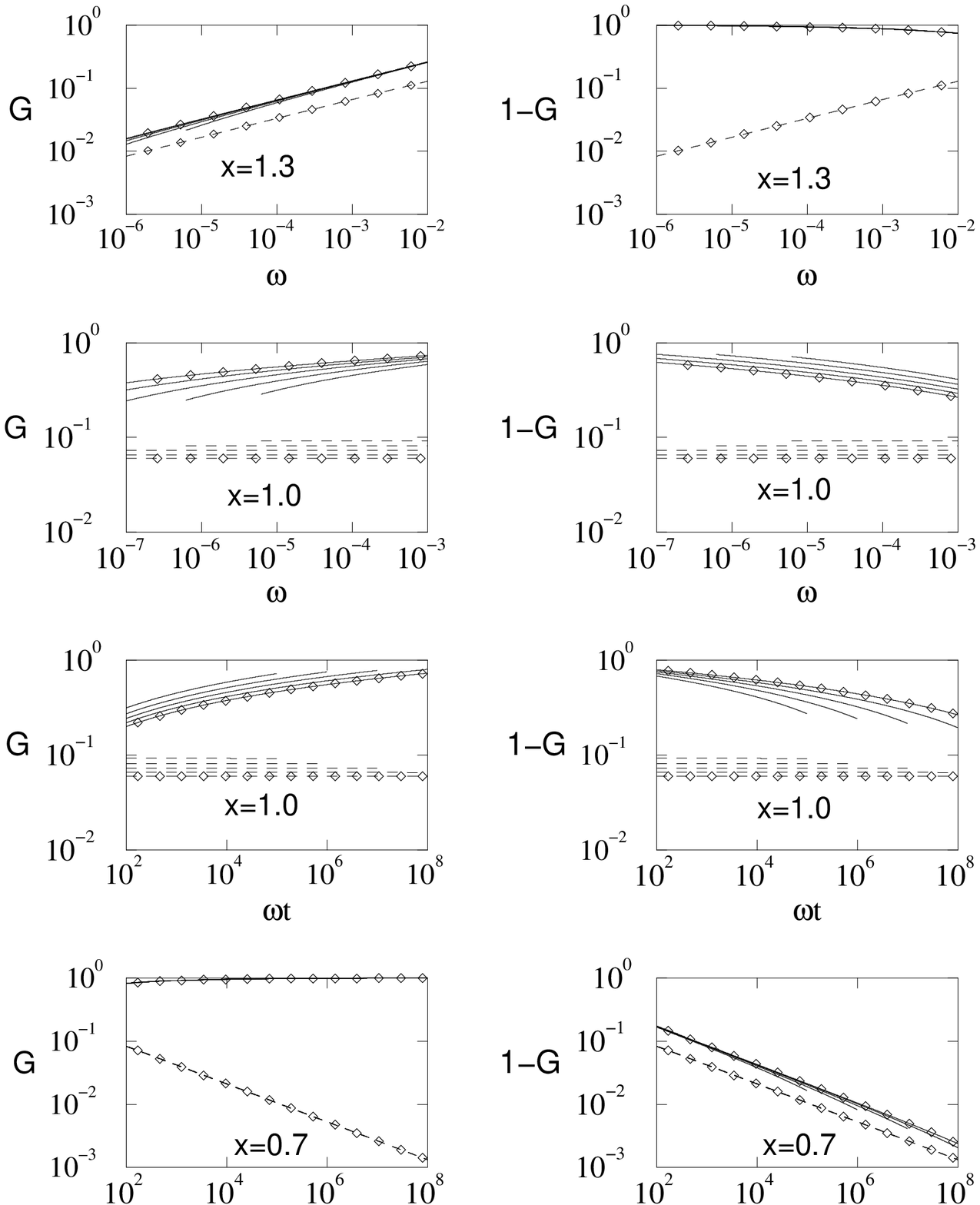,width=14cm}
\end{center}
\caption{Left column: viscoelastic spectra $G'(\omega)$ (solid lines) and
$G''(\omega)$ (dashed lines) versus frequency, $\omega$ (for $x\ge 1$) or
scaled frequency $\omega
t$ (for $x\le 1$). Right column: frequency-dependent corrections to Hookean
elasticity, $1-G'$
(solid lines), $G''$ (dashed lines). Data are shown for systems aged
$t=10^7\,, 10^8\,, \ldots
10^{11}$. At any fixed $\omega$ the curves lie in order of age; data on the
oldest system is marked
by the symbols. There is good data collapse both above and below the glass
point (with the
appropriate scalings); the data for $x=1$ do not collapse in either
representation due to
logarithmic terms. If plotted against $\omega$ rather than $\omega t$,
the data for $x=0.7$ would resemble Fig.~\ref{fig:possible_modes}. 
\label{fig:G_osc_time_dep} }
\end{figure}

Numerical results for the viscoelastic spectrum $G^*(\omega,t)$ at various
measurement times $t$
for various $x$ are shown in Fig.\ref{fig:G_osc_time_dep}. These indeed
show a characteristic
``hardening" of the glassy material as it ages: the storage modulus at low
frequencies evolves
upwards, and the loss modulus downward~\cite{SolLeqHebCat97,long_El}.  Each
spectrum terminates at
frequencies of order $\omega t
\simeq 1$. This is because one cannot measure a true oscillatory response
for periods {\em beyond}
the age of the system\footnote{And, although~\(eqn:G_osc_lin) still
provides an unambiguous
definition of $G^*(\omega,t,t_{\rm s})$, this ceases to be independent
of $t_{\rm s}$ in this regime, so $G^*(\omega,t)$ is
undefined.}. Therefore, the rise at low frequencies in $G'$ spectra
like Fig.~\ref{fig:possible_modes} represents the ultimate rheological
behavior\footnote{Note that this only applies for $\mu = 1$ in
Struik's scheme, as exemplified by SGR. Whenever $\mu < 1$, the region
to the left of the loss peak can, in principle, be accessed
eventually.}.

It is shown in App.~\ref{app:ts_indep} that the insensitivity of
$G^*(\omega,t,t_{\rm s})$ to
$t_{\rm s}$ in practical measurements of the viscoelastic spectrum (where an
oscillatory strain is
maintained over many cycles) arises because (even when $x<1$) the most
recently executed strain
cycles dominate the stress response at time $t$. In essence, the result
means that, as long as
oscillatory strain was started many cycles ago, there is no memory of when
it was switched on;
accordingly (by linearity) an oscillatory strain started in the distant
past and then switched off
at $t_{\rm s}$, will leave a stress that decays on a timescale
comparable to the 
period of the
oscillation. This is markedly different from non-oscillatory stresses,
where long term memory
implies that the response to 
a step strain, applied for a long time, persists for
a similarly long time
after it is removed  (see Sec.~\ref{sec:ageing} above). Thus the fact that
the SGR glass ``forgets"
the $t_{\rm s}$ argument of $G^*(\omega,t,t_{\rm s})$, is directly
linked to the 
oscillatory nature of the
perturbation.
As also shown in App.~\ref{app:ts_indep}, this forgetfulness means that, in
the SGR model, a
Fourier relationship between oscillatory and step strain responses is
recovered; to a good
approximation, one has the relation~\(eqn:maybe)
\begin{equation} G^*(\omega,t) = i\omega \int_0^\infty e^{-i\omega t'}
G(t',t)\, dt'
\label{eqn:G_star_is_ft_two}
\end{equation}
 Apart from the explicit dependence on the measurement
time\footnote{Formally, $t$
appears as the time at which an step strain was initiated, or an
oscillatory measurement ended.
Thus $G^*(\omega,t)$ is to within $i\omega$, the Fourier transform of the
step strain response
function
$G(\Delta t,t)$ that would be measured if a step strain were applied
immediately {\em after} the
oscillatory measurement had been done.} $t$, this is the usual (TTI)
result. But here the
result is nontrivial due to the presence of ageing effects. As
discussed at the end of Sec.~\ref{sec:ageing}, we speculate that the
relation~\(eqn:G_star_is_ft_two) holds not only for the SGR model, but
in fact for all systems which have only {\em weak} long term memory.

\subsubsection{Startup of Steady Shear}
\label{sec:shear_start_up_linear}

Consider now a startup experiment in which a steady shear of rate
$\gdot \ll 1$ is commenced at time $t_{\rm w}$. So long
as we restrict attention to times short enough that the total strain
remains small ($\gdot(t-t_{\rm w})
\ll 1$) the system remains in a linear response regime.\footnote{This
contrasts with the ultimate
steady-state behaviour which, for $x<2$, is always nonlinear; the crossover
to a nonlinear regime at
late times is discussed in Sec.~\ref{sec:steady_shear_nonlin} below.}

Within the regime of linear response, any element's lifetime is independent
of strain and obeys
$\tau = \exp(E/x)$. As described in Sec.~\ref{sec:glass_transition}
above, at a time $t$ after a
deep quench, the distribution of lifetimes obeys
$P(\tau,t)\sim\tau\rho(\tau)$ for $\tau \ll t$ and
$P(\tau,t)\sim t \rho(\tau)$ for $\tau \gg t$. Since the local stress
associated with a given trap
is of order $\gdot \tau$ for $\tau \ll t-t_{\rm w}$, and
$\gdot (t-t_{\rm w})$ for $\tau \gg t-t_{\rm w}$,
we can construct an estimate of the macroscopic
stress; for
$t-t_{\rm w}
\ll t_{\rm w}$,
\begin{eqnarray}
\label{eqn:stress_evolution_approx}
\sigma(t)&\simeq &\frac{\gdot\left[\int_1^{t-t_{\rm w}}  \tau^2
\rho(\tau) d\tau 
+ (t-t_{\rm w})\int_{t-t_{\rm w}}^t
 \tau \rho(\tau) d\tau+ (t-t_{\rm w})t \int_t^\infty
  \rho(\tau)d\tau\right]}{\int_1^t \tau\rho(\tau)d\tau +
 t\int_t^\infty \rho(\tau)d\tau}\nonumber\\
        \, &\simeq & \frac{\gdot\left[ x(t-t_{\rm w}) ^{2-x}+
(x-2)(t-t_{\rm w})t^{1-x}+x(1-x)\right]}{(x-2)\left(t^{1-x}-x\right)}
\end{eqnarray}
 This gives, for long times and in the linear response
regime, $\sigma(t) \sim
\dot\gamma(t-t_{\rm w})$ for $x<1$ (which is purely elastic behaviour),
$\sigma(t) \sim \dot\gamma (t-t_{\rm w})^{2-x}$ for $1<x<2$ (which is an
anomalous power law), and
$\sigma(x) \sim \gdot $  for $x>2$; repeating the same calculation with
$t\gg t_{\rm w}$ gives the same
asymptotic scaling in each case. An asymptotic analysis of the constitutive
equations confirms
these scalings, with the prefactors as summarized in
table~\ref{table:startup}. Because the results depend only on
$t-t_{\rm w}$, any explicit dependence on $t_{\rm w}$
(ageing, or anomalously slow transients) must reside in
subdominant corrections
to these leading asymptotes.  Accordingly, {\em linear} shear startup is
not a good experimental
test of such effects (but see Sec.~\ref{sec:steady_shear_nonlin} below).
\renewcommand{\arraystretch}{2.6}
\begin{table}[tb!]
\begin{center}
\begin{tabular}{|l||c|c|}\hline
        & $\D \frac{\sigma(t-t_{\rm w},t_{\rm w})}{\gdot}\ \
\mbox{for}\ \ t-t_{\rm w}\ll t_{\rm w}$ 
        & $\D \frac{\sigma(t-t_{\rm w},t_{\rm w})}{\gdot}\ \
\mbox{for}\ \ t-t_{\rm w}\gg t_{\rm w}$ 
          \\[5mm]\hline\hline
$2<x$ & $\D \frac{x-1}{x-2}$
        & $\D \frac{x-1}{x-2}$
          \\[5mm] \hline
$1<x<2$ & $\D \frac{\Gamma(x)}{2-x}(t-t_{\rm w})^{2-x}$
        & $\D \frac{(x-1)\Gamma(x)}{2-x}(t-t_{\rm w})^{2-x}$
          \\[5mm] \hline
$x<1$   & $\D (t-t_{\rm w})$
        & $\D (1-x)(t-t_{\rm w})$
          \\[5mm] \hline
\end{tabular}
\caption{Stress response to shear strain of constant rate $\gdot$  at short
and long times
($t-t_{\rm w}\gg 1$, $t\gg 1$, $\gdot\ll 1$ assumed). These results apply to
the regime $\gdot(t-t_{\rm w})\ll 1$, where strain-induced yielding can be
neglected, making the response linear.
$\Gamma(x)$ denotes the Gamma function.
\label{table:startup}
}
\end{center}
\end{table}
\renewcommand{\arraystretch}{1}
The power law anomaly for $1<x<2$ can be understood by examining
which traps make dominant contributions to
$\sigma(t) = \int s(\tau,t) d\tau$. (Recall that $s(\tau,t)d\tau$ is the
stress contribution at
time $t$ from elements of lifetime $\tau$; see
Sec.~\ref{sec:coneq}.) For $x>2$,
$s(\tau,t)$ is weighted strongly toward traps of lifetime
$O(1)$; hence $\sigma(t)$ tends to a finite limit (of order $\gdot$) as
$t\rightarrow\infty$, and the viscosity of the system is finite. For
$x<2$, on the other hand, most of the weight in the
$s(\tau,t)$ distribution involves lifetimes of order $t$. As time passes,
stress is carried by
deeper and deeper traps, and (in the absence of flow-induced yielding) the
mean stress diverges as
$t\rightarrow\infty$.

In fact, as discussed in Sec.~\ref{sec:higher_transition} above, just as
the Boltzmann distribution
for the relaxation times
$P_{\rm eq}(\tau) = P(\tau,
\infty)
\sim\tau\rho(\tau)$  is non-normalisable for
$x\le 1$ (giving glassiness and ageing), so, in the absence of
strain-induced yielding, is the
ultimate distribution
$s(\tau,\infty)\sim \tau^2\rho(\tau)$ of stresses residing in traps of
lifetime $\tau$, whenever
$x<2$. The zero shear viscosity $\eta$ is therefore infinite throughout
this regime, as noted
previously.

\subsection{Nonlinear Response}

We now turn to the nonlinear behaviour of the SGR model under imposed
strain, starting with the
step strain case.

\subsubsection{Step Strain}
\label{sec:results_strain_nonlin_step} The nonlinear step strain response
function was defined in~\(eqn:steplin). It is found for the SGR model
from~\(eqn:ceone_alt):
\begin{equation}
\label{eqn:nonlin_step} G(t-t_{\rm w},t_{\rm w};\gamma_0)=G_{0}
(Z(t,0))+\int_0^{t_{\rm w}} 
Y(t') G_{\rho}
(Z(t,t')) dt'
\end{equation}
 where, using~\(eqn:Z):
\begin{equation}
\label{eqn:Z_for_step}
Z(t,t')=(t-t_{\rm w})\exp\left(\gamma_0^2/2x\right) + (t_{\rm w}-t')
\end{equation} On the other hand, in the linear regime we have:
\begin{eqnarray}
\label{eqn:lin_step_recap} G(t-t_{\rm w},t_{\rm w},\gamma_0\to 0)&\equiv&
G(t-t_{\rm w},t_{\rm w})\nonumber\\
                          &=& G_{0} [(t-t_{\rm w})+(t_{\rm w}-0)]\nonumber\\
                          & &  + \int_0^{t_{\rm w}} Y(t')
G_{\rho}\left[(t-t_{\rm w})+(t_{\rm w}-t')\right]  dt'
\end{eqnarray} Direct comparison of~(\ref{eqn:nonlin_step})
and~(\ref{eqn:lin_step_recap}) reveals that:
\begin{equation}
\label{eqn:nonlin_step_final}
G(t-t_{\rm w},t_{\rm w};\gamma_0)=G\left((t-t_{\rm
w})\exp\left(\gamma_0^2/2x\right),t_{\rm w}\right) 
\end{equation} 
This result generalizes that of~\citet{long_El} for the non-ageing case
($x>1$). It can be understood as follows. Within the SGR model,
instantaneous response to a step strain at $t_{\rm w}$ is always
elastic (that is, $G(0,t_{\rm w},\gamma_0) = 1$); the fraction of
stress remaining at time $t>t_{\rm w}$ is the fraction of elements
which have survived from $t_{\rm w}$ to $t$ without yielding (see
Sec.~\ref{sec:results_strain_linear_step} above). The stress decay is
therefore determined entirely by the distribution of relaxation times
in the system just after the strain is applied at time $t_{\rm
w}$. The effect of a finite strain is solely to modify the
distribution of barrier heights, and hence to modify this distribution
of relaxation times $\tau$; in fact (within the model) nonlinear
strain reduces the yield time of every element by an identical factor
of $\exp(\gamma_0^2/2x)$~\cite{long_El}. Thus the relaxation after a
nonlinear step strain at $t_{\rm w}$ is found from the linear case by
rescaling the time interval $t-t_{\rm w}$ using this same
factor. Accordingly, the asymptotic results given for $G(t-t_{\rm
w},t_{\rm w})$ in table~\ref{table:modulus} can be converted to those
for the nonlinear regime by replacing the time interval $t-t_{\rm w}$
by a strain-enhanced value $(t-t_{\rm w})\exp(\gamma_0^2/2x)$,
wherever it appears there.

\subsubsection{Startup of Steady Shear}
\label{sec:steady_shear_nonlin}

In Sec.~\ref{sec:shear_start_up_linear} we discussed the response to
start up of steady shear (with $\gdot \ll 1$) at time $t_{\rm w}$; we
assumed there that a linear response was maintained. Let us now
consider the effect of strain-induced yield events, which cause
nonlinearity. Consider first what happens for $x>2$ (where the SGR
model predicts Newtonian fluid behaviour for $\gdot\ll 1$). Here the
main stress contribution is from elements which, were they unstrained,
would have lifetime $\tau(E)=\exp(E/x)$ of order unity. So, if the
strain rate obeys $\gdot\ll 1$, these elements will acquire only
negligible stress before they yield spontaneously.  Hence their
lifetimes are not affected by strain, and the stress response remains
linear at all times, including the steady state limit:
$\sigma(t\rightarrow\infty) \to \eta \gdot$.

In the following, we focus on the case $x<2$, where nonlinearities do
appear. The dominant stress contributions in this noise temperature
regime are from deep
traps, {\em i.e.}, elements with lifetimes of order
$t$. Linearity applies only if such elements are unlikely to undergo
strain-induced yielding before they yield spontaneously, after a time
of order $t$.
Such elements carry strains
of order $\gdot t$, which enhances their yield rate by a factor $\exp[(\gdot
t)^2/2x]$; we require
that this is small, which holds only so long as
$\gdot t \ll 1$. Hence the predictions of the linear theory of
Sec.~\ref{sec:shear_start_up_linear}  can be maintained to arbitrarily long
times only by taking
the limit
$\gdot \to 0$ before one takes the steady state limit of $t\to\infty$. This
means that the width of the 
linear response regime in steady flow is vanishingly small for $x<2$,
as previously discussed.

As mentioned in Sec.~\ref{sec:shear_start_up_linear}, throughout the linear
period the startup
curve shows no strong ageing or transient effects, even though the stress
is evolving into deeper
traps.  At finite $\gdot$, the linear period ends at $t\simeq \gdot^{-1}$
(within logarithmic
terms, discussed below); at later times, the main stress-bearing elements
will, during their
lifetimes, become strongly strained. Indeed, at strain rate $\gdot$,
an element with yield energy $E$ will be strained to the top of its
yield barrier in a time $t_{\rm int}
 \simeq E^{1/2}/\gdot \simeq
(\log\tau)^{1/2}/\gdot$.
 The tendency of the stress distribution $s(\tau,t)$ (and
also, for any
$x<1$, the lifetime distribution $P(\tau,t)$) to evolve toward deeper and
deeper traps is thereby
{\em interrupted}: the lifetime of a deep trap is converted from $\tau$ to
a much smaller value, of
order $(\log\tau)^{1/2}/\gdot$~\cite{SolLeqHebCat97,long_El}.  This
truncation of the lifetime
distribution is enough to ensure that these distributions are never
dominated by the deep traps,
and a steady state is recovered; accordingly, there are no ageing effects
at late enough times
either.

\begin{figure}[h]
\begin{center}
\epsfig{file=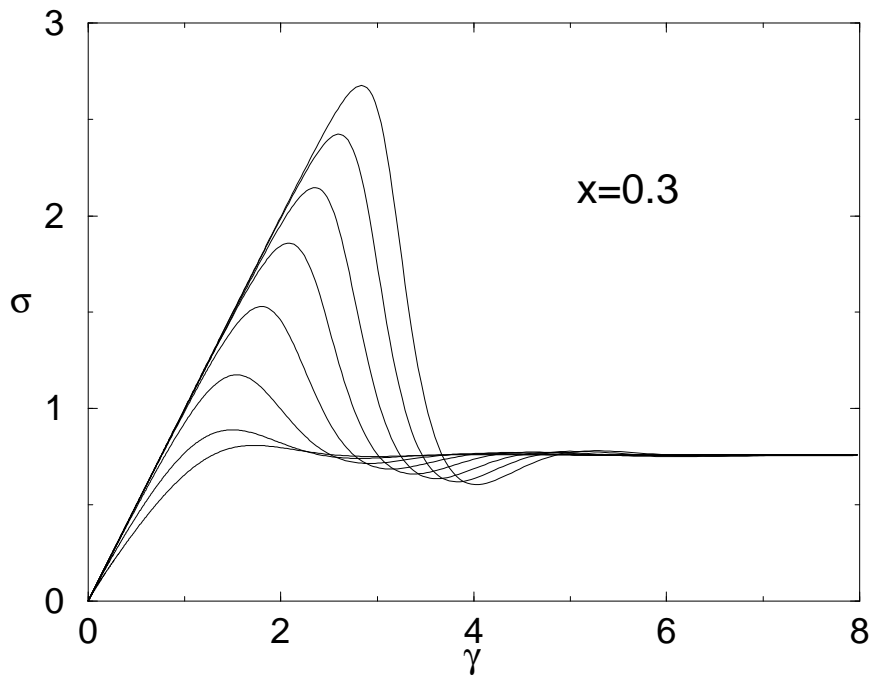,width=11.5cm}
\end{center}
\caption{Stress response $\sigma$, in shear startup, vs strain $\gamma$ at
noise temperature
$x=0.3$ and strain rate $\gdot=0.001$. Curves from bottom to top correspond
to increasing ages
$t_{\rm w}=10^2\,,10^3 \ldots\, 10^9$ at time of startup.
\label{fig:startup} }
\end{figure}

Note, however, that the stress at the end of the linear regime can be
higher than the
steady state value, leading to an overshoot in the startup curve; 
see~\citet{long_El}.  This overshoot region, unlike the two
asymptotes, shows a
significant dependence on the system age $t_{\rm w}$, as shown in
Fig.~\ref{fig:startup}. The physics of
this is clear: the extent of the linear regime progressively gets larger as
$t_{\rm w}$ is increased, because the system has aged into deeper
traps (and because the SGR model assumes that within each trap the
relation between stress and strain is linear). Thus the
strain at which strong yielding sets in increases (roughly logarithmically)
with
$t_{\rm w}$; the height of the overshoot is accordingly increased
before dropping onto the same,
$t_{\rm w}$-independent, steady-shear plateau.

\section{Rheological Ageing: Imposed Stress}
\label{sec:results_stress}

We now analyse the SGR model's predictions for various stress-controlled
rheological experiments.
(We continue to assume the sample to have been prepared at time $t=0$ by
the  idealized ``deep
quench'' procedure defined in Sec.~\ref{sec:sample_preparation}.) As
previously remarked, the
structure of the  constitutive equations makes the analysis more difficult
for imposed stress than
for imposed strain. The following discussion is therefore largely based on
our numerical results,
with asymptotic analysis of a few limiting cases. Our numerical method is
outlined in
App.~\ref{app:iterative_code}.

\subsection{Linear Response}
\label{sec:results_stress_linear}

\subsubsection{Step Stress}
\label{sec:results_stress_linear_step}

The SGR model predicts that upon the application of a step stress there
will be an instantaneously
elastic response. Elements then progressively yield and reset their local
stresses to zero; thus we
must apply progressively more strain to maintain the macroscopic stress at
a constant value. In
this way strain is expected to increase with time (but at a rate that could
tend to zero at long
times). Potentially therefore, individual elements can acquire large local
strains and, just as in
the shear startup case, linearity of the response need not be maintained at
late times. As we did
for shear startup, we therefore 
first proceed by assuming that the response {\em is}
linear; we find the
corresponding
$\gamma(t)$ and then (in Sec.~\ref{sec:results_stress_nonlin}  below)
consider {\em a posteriori}
up to what time $t$ the linear results remain valid.

In the linear regime the step stress response is described by the creep
compliance $J(t-t_{\rm w},t_{\rm w})$
which was defined for non TTI systems in Sec.~\ref{sec:creep}. We
computed this quantity
numerically from the linearized form of the constitutive
equation~\(eqn:modulus_from_ce) for the SGR model,
which for step stress may
be written
\begin{equation} 1=J(t-t_{\rm w},t_{\rm w})-\int_{t_{\rm w}}^t
J(t'-t_{\rm w},t_{\rm w})Y(t') 
G_{\rho}(t-t') dt'
\label{eqn:con_J}
\end{equation}
In analysing our numerical results we first identify, as
usual, regimes of short and
long time interval between stress onset and measurement,
$t-t_{\rm w} \ll t_{\rm w}$ and $t-t_{\rm w} \gg t_{\rm w}$
respectively.\footnote{As before, 
we apply the ``macroscopic
time'' conditions $t-t_{\rm w} \gg 1$ and $t_{\rm w}\gg 1$.} In these
two regimes we 
find the time
dependences summarized in table~\ref{table:compliance}.  For the long time
interval regime ($t-t_{\rm w}
\gg t_{\rm w}$), the results were in fact obtained as follows. Curves for
$J(t-t_{\rm w},t_{\rm w})$ were first generated numerically; the
observed scalings (for 
example,
$J\sim (t-t_{\rm w})^{x-1}$ for $1<x<2$) were then taken as ans\"atze for
analytic substitution into the
constitutive equation~\(eqn:con_J). In each case this allowed us to confirm
the given functional
form, and to compute exactly the  $x$-dependent prefactors shown. These
prefactors were
cross-checked by comparison with the numerical results; no discrepancies
were found within available
accuracy.

\renewcommand{\arraystretch}{2.7}

\begin{table}[htb!]
\begin{center}
\begin{tabular}{|l||c|c|}\hline
        & $\D J(t-t_{\rm w},t_{\rm w})\ \mbox{for}\ t-t_{\rm w}\ll t_{\rm w}$
        & $\D J(t-t_{\rm w},t_{\rm w})\ \mbox{for}\ t-t_{\rm w}\gg t_{\rm w}$
          \\[5mm]\hline\hline
$x>2$   & $\D \frac{x-2}{x-1}\,(t-t_{\rm w})$
        & $\D \frac{x-2}{x-1}\,(t-t_{\rm w})$ 
          \\[5mm]\hline
$1<x<2$ & $\D \frac{(t-t_{\rm w})^{x-1}}{\Gamma^2(x)\Gamma(2-x)}$
        & $\D \frac{(t-t_{\rm w})^{x-1}}{\Gamma^2(x)\Gamma(2-x)-\Gamma(x)}$
          \\[5mm] \hline
$x=1$   &  ---
        & $\D \frac{3}{\pi^2}\ln^2(t-t_{\rm w})$  \\ \hline
$x<1$   &  ---
        & $\D \frac{1}{\psi(1)-\psi(x)}\ln\left(\frac{t-t_{\rm
w}}{t_{\rm w}}\right)$ 
          \\[5mm] \hline
\end{tabular}
\caption{Linear creep compliance in the SGR model at long and short times
  ($t-t_{\rm w}\gg 1$, $t_{\rm w}\gg 1$ assumed). $\Gamma(x)$ denotes
  the Gamma function, and $\psi(x)=\Gamma'(x)/\Gamma(x)$
the usual psi function. The blank entries for $x<1$ are explained in
the text; see also footnote~\ref{Jshort_x_lt_one}.
\label{table:compliance}
}
\end{center}
\end{table}

\renewcommand{\arraystretch}{1}

To obtain results for short time intervals, we proceeded by assuming that
the resulting compliance
$J(t-t_{\rm w},t_{\rm w})$ is the same as if we first let $t_{\rm w}
\to \infty$ (the 
dominant traps are in Boltzmann
equilibrium; see Fig.~\ref{fig:prob_distr}a); this limits\footnote{For $x
<1$, we find instead $J=1+\mbox{const}\times\left[(t-t_{\rm w})/t_{\rm
w}\right]^{1-x}$ 
at very early times; but this breaks down
as soon as the second
term becomes comparable to the leading (elastic) result.
\label{Jshort_x_lt_one}} the analysis to
$x>1$. The resulting  prediction of
$J(t-t_{\rm w},t_{\rm w}\to\infty)$ was found analytically from
$G(t-t_{\rm w},t_{\rm w}\to\infty)$ and the reciprocal relations between the
corresponding Fourier transforms (see Sec.~\ref{sec:spectra} above);
these were again checked numerically.

\begin{figure}[h]
\begin{center}
\epsfig{file=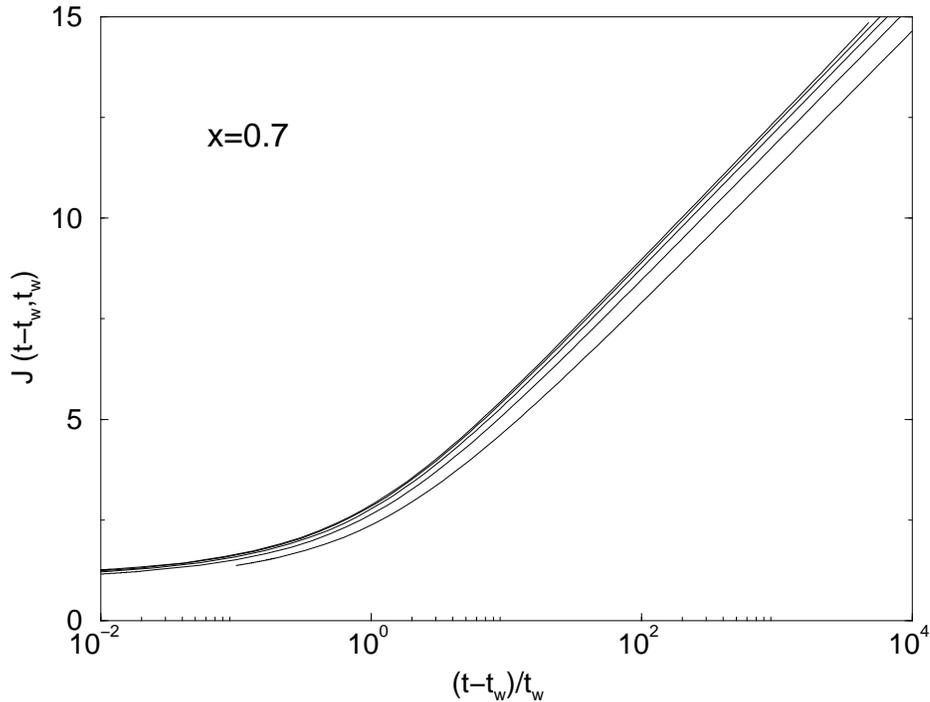,width=11.5cm}
\end{center}
\caption{Linear creep compliance $J(t-t_{\rm w},t_{\rm w})$ against scaled time
interval $(t-t_{\rm w})/t_{\rm w}$ for
noise temperature $x=0.7$. Curves from bottom to top correspond to
increasing times
$t_{\rm w}=10^2\,,10^3 \ldots 10^6$ of stress onset. Note the approach to a limiting
scaling form as $t_{\rm w}$
becomes very large compared with the microscopic time $\tau_0=1$.
\label{fig:step_stress_linear_x0.7} }
\end{figure}

\begin{figure}[h]
\begin{center}
\epsfig{file=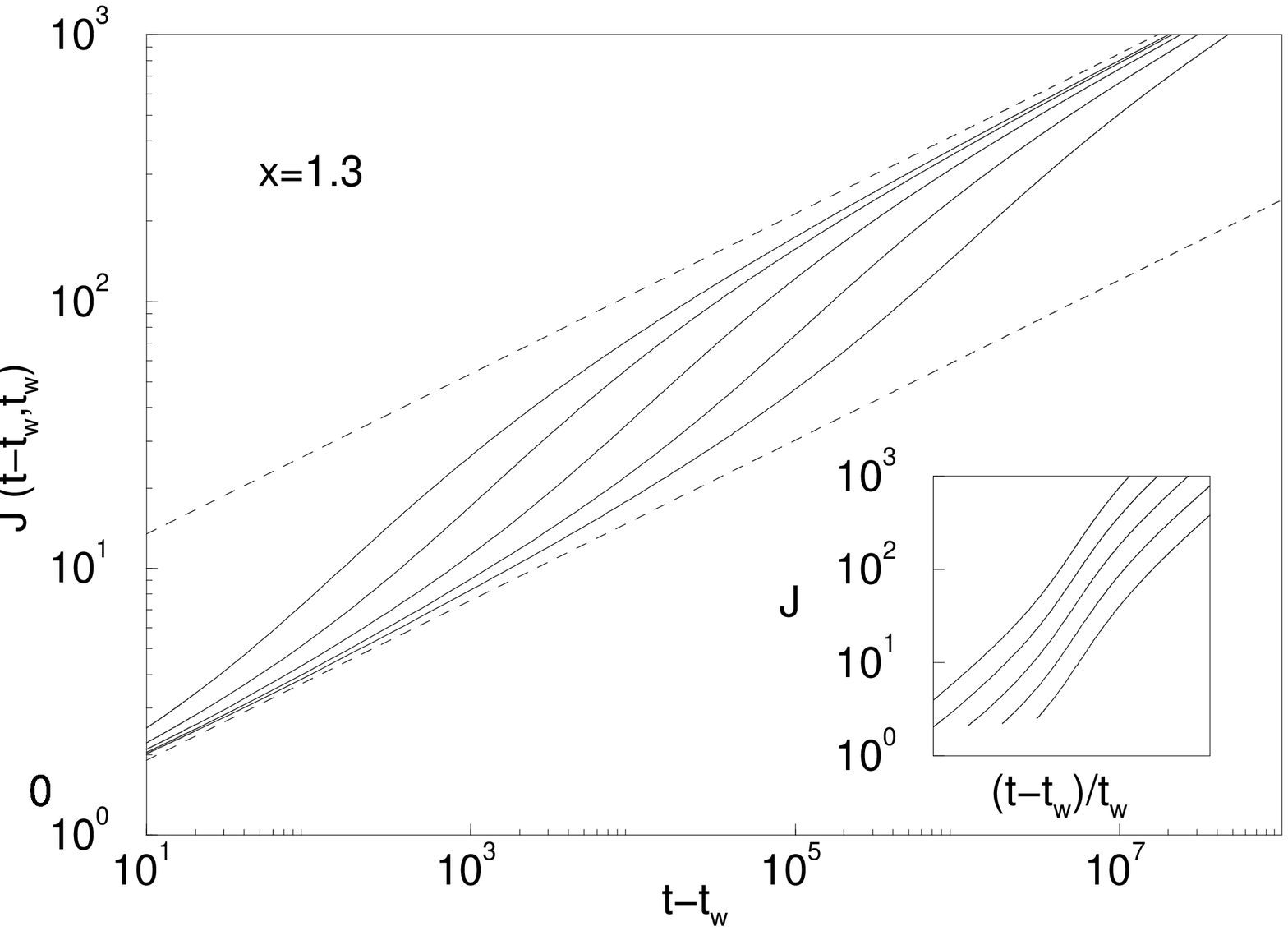,width=11.5cm}
\end{center}
\caption{Linear creep compliance $J(t-t_{\rm w},t_{\rm w})$ against
time interval $t-t_{\rm w}$ for noise temperature $x=1.3$ for $t_{\rm
w}=10^3\,,10^4 \ldots 10^7$ (solid lines, top to bottom). Lower dashed
line: theoretical prediction for the short time regime $t-t_{\rm w}\ll
t_{\rm w}$. Upper dashed line: asymptote for long time regime
$t-t_{\rm w}\gg t_{\rm w}$. Note that short time and long time
behaviours are each independent of $t_{\rm w}$ (as expected for
$x>1$), but that the crossover time between them scales with $t_{\rm
w}$. Inset: same data plotted against scaled time $(t-t_{\rm
w})/t_{\rm w}$; the order of the curves is reversed. The crossover
between short and long time behaviour takes place at a value of
$(t-t_{\rm w})/t_{\rm w}$ which is roughly the same for all curves,
demonstrating the scaling with $t_{\rm w}$.
\label{fig:step_stress_linear_x1.3}
}
\end{figure}

Further insight into the results of table~\ref{table:compliance} can be
gained as follows. In
step stress, we need to keep applying larger and larger strains because
elements progressively
yield and reset their local stresses to zero. To maintain constant stress,
the rate at which stress
increases due to straining, which in our units is just the strain rate
$\gdot$, must match the rate at which stress is lost, due to local yielding
events. The latter
defines a ``stress-weighted hopping rate"
$Y_s=\int
\tau^{-1}s(\tau,t) d\tau$. For $x>2$, $Y_s$ remains a constant of order
$\sigma_0$; stress remains
in traps of lifetime $\tau=O(1)$ and the creep response is purely viscous.
For $x<2$, however,
$Y_s$ decays as a power law\footnote{In fact, $Y_s\sim\gdot\sim
(t-t_{\rm w})^y$ 
where $y=x-2$ for
$1<x<2$ and $y=-1$ for $x<1$.} of
$(t-t_{\rm w})$; the stress distribution $s(\tau,t)$ is dominated at
time $t$ by 
traps with lifetimes
$\tau$ of order $t-t_{\rm w}$, the time interval since the stress application.

For $1<x<2$, the scenario given above for the time-dependence of $Y_s$
is closely analogous to that given in Sec.~\ref{sec:glass_transition}
above for the hopping rate $Y =\int \tau^{-1} P(\tau,t)d\tau$ in
systems with $x<1$. Indeed, the evolution of $Y_s$ following a step
stress, at noise temperature $x$, is closely related\footnote{More
  generally one can show for the SGR model that, for an {\em
    equilibrium} system whose noise temperature is $x>1$, the
  evolution of the stress distribution $s(\tau,t)$ following
  application of a step stress at $t=0$ is, at long times,
  equivalent to that of the probability distribution
  $P(\tau,t)$, in a system deep-quenched to a noise temperature $x-1$
  at $t=0$. This result is connected with the discussion made
  in Sec.~\ref{sec:higher_transition} above, of the variation with $x$
  of the dynamics of successive moments of the lifetime distribution:
  at noise temperature $x+n$, the dynamics of the $n$th moment is like
  that of the zeroth moment at noise temperature $x$.}  to that of
$Y$, following a quench, at noise temperature $x-1$.

The ageing behaviour of the linear creep compliance $J(t-t_{\rm
w},t_{\rm w})$ shows significant differences from the step strain
modulus $G(t-t_{\rm w},t_{\rm w})$ discussed in
Sec.~\ref{sec:results_strain_linear_step} above\footnote{Since it
refers to a shear measurement, one would not expect our result to
resemble the empirical (stretched exponential) form measured
by~\citet{Struik78} for a wide range of materials in {\em tensile}
creep; nonetheless, it shows upward curvature on a log-log plot before
approaching the eventual logarithmic form (with downward
curvature). The same applies in nonlinear creep; see
Fig.~\ref{fig:step_stress_below_glass} below.}.  In the glass phase
($x<1$), the strain response to step stress indeed depends on age: it
is a function of $(t-t_{\rm w})/t_{\rm w}$ as expected (see
Fig.~\ref{fig:step_stress_linear_x0.7}).  However, the dependence (for
long time intervals) is only logarithmic; $J(t-t_{\rm w},t_{\rm w})
\sim \ln\left((t-t_{\rm w})/t_{\rm w}\right) = \ln(t-t_{\rm w}) -\ln
t_{\rm w}$ (see table~\ref{table:compliance}) which means that in the
long time interval limit ($t-t_{\rm w} \gg t_{\rm w}$) the explicit
waiting time dependence ($\ln t_{\rm w}$) represents formally a
``small" correction to the leading behaviour $\ln(t-t_{\rm w})$. This
relatively slight $t_{\rm w}$-dependence in creep measurements is
intuitively reasonable: the strain response at time $t$ to step stress
is {\em not} determined purely by the relaxation spectrum at $t_{\rm
w}$ (as was the case in step strain, table~\ref{table:modulus}), but
by the dynamics of the system over the entire interval between $t_{\rm
w}$ and $t$. This decreases the sensitivity to the time $t_{\rm w}$ at
which the perturbation was switched on. Similar remarks hold above the
glass point ($1<x<2$, see Fig.~\ref{fig:step_stress_linear_x1.3}): in
step strain, we found for $t-t_{\rm w} \gg t_{\rm w}$ a slow transient
behaviour which depended to leading order upon $t_{\rm w}$
(table~\ref{table:modulus}). For step stress, however, the
corresponding $t_{\rm w}$ dependence is demoted to lower order, and
the late-time response is dominated by TTI terms.\footnote{We restate
here why we call these effects for $x>1$ transient behaviour rather
than ageing. As explained after eq.~\(WLTM), a consistent definition
of long term memory and ageing for the step stress response function
$J(t-t_{\rm w},t_{\rm w})$ requires a form of ``regularization'' by
considering the material in question in parallel with a spring of
infinitesimal modulus $g$. This effectively puts an upper limit of
$J_{\rm max}=1/g$ on the observable values of $J(t-t_{\rm w},t_{\rm
w})$.  Taking the limit $t_{\rm w}\to\infty$ for $x>1$ then results in
a fully TTI step stress response, whatever the value of $J_{\rm
max}$. On the other hand, for $x<1$, the (albeit weak, logarithmic)
$t_{\rm w}$-dependence of the response remains visible even for finite
values of $J<J_{\rm max}$.  }

\subsubsection{Oscillatory Stress}
\label{sec:results_stress_linear_osc}

For the SGR model it was noted in
Sec.~\ref{sec:results_strain_linear_osc} that ($i$) in the oscillatory
stress response
$G^*(\omega,t,t_{\rm s})$, the $t_{\rm s}$ dependence is negligible
for low frequencies 
($\omega \ll 1$)
whenever
$\omega(t-t_{\rm s}) \gg 1$ and $\omega t_{\rm s} \gg 1$; ($ii$) these
conditions 
are satisfied in most
conventional rheometrical measurements of the viscoelastic spectrum, where
an oscillatory strain is
maintained for many cycles;  and ($iii$), perhaps surprisingly, these
facts are true even in
the glass phase, $x \le 1$, of the SGR model.  We also noted that, because
response to oscillatory
strain is dominated by memory of the few most recent cycles (over which the
system has barely aged),
$G^*(\omega,t)$ is the Fourier transform (with respect to the time interval
$\Delta t$) of the
step strain response function $G(\Delta t,t)$ that would be measured if a
step strain were applied
immediately after the oscillatory measurement  had been
done~\(eqn:G_star_is_ft_two).

\begin{figure}[h]
\begin{center}
\epsfig{file=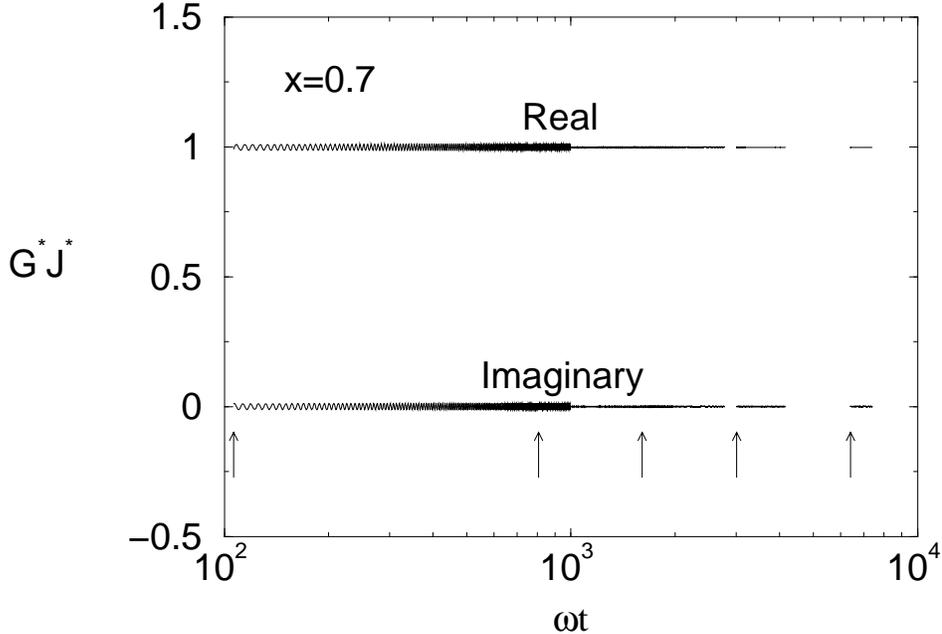,width=11.5cm}
\end{center}
\caption{Real and imaginary parts of the product
  $G^*(\omega,t)J^*(\omega,t)$ vs $\omega t$ at noise temperature
  $x=0.7$ and frequency $\omega=0.01$. The usual reciprocity relation
  between $G^*$ and $J^*$ is seen to hold to within about one percent.
  Shown are the results of several runs, each over a different time
  window.  A vertical arrow marks the horizontal co-ordinate of the
  start of each data set.  In each run shearing was commenced 20
  cycles before the start of data output, to ensure that
  $\omega(t-t_{\rm s})\gg1$ (necessary for $t_{\rm s}$-independence).
  The oscillatory deviations, visible for the leftmost data set, arise
  because the other condition for $t_{\rm s}$-independence ($\omega
  t_{\rm s} \gg 1$) is only just satisfied.
\label{fig:reciprocity} }
\end{figure}

We have confirmed numerically that similar remarks apply to the oscillatory
stress response function
$J^*(\omega,t,t_{\rm s})$.\footnote{Although unsurprising, this does require
explicit confirmation since, for example, the transient effects from
switching on the perturbation
could be different in the two cases.} This was defined in
Sec.~\ref{sec:spectra_nontti} as the
strain response, measured at
$t$, to an oscillatory stress initiated at time $t_{\rm s}$.  Memory of the
startup time $t_{\rm s}$ is indeed small in
$J^*(\omega,t,t_{\rm s})$ so long as
$\omega(t-t_{\rm s}) \gg 1, \omega t_{\rm s} \gg 1$ (and $\omega \ll
1$). It appears 
that, just as in
the case of a strain controlled experiment, strain response to oscillatory
stress is dominated by
memory to the most recent cycles, over which the system has barely aged. We
may therefore suppress
the $t_{\rm s}$ parameter, defining a compliance spectrum at time
$t$ by $J^*(\omega,t)$. Furthermore, $J^*(\omega,t)$ is found numerically
to be the reciprocal of
$G^*(\omega,t)$,
\begin{equation} J^*(\omega,t)G^*(\omega,t)=1
\end{equation} just as it is (without the $t$ argument) in normal TTI
systems. The numerical
confirmation of this result is presented in Fig.~\ref{fig:reciprocity}. We
emphasize that this
result, like the previous one, has been confirmed here specifically for the
SGR model; but it may hold more widely for systems with weak long term
memory (see Sec.~\ref{sec:ageing}).

\subsection{Nonlinear Response}
\label{sec:results_stress_nonlin}

\subsubsection{Step Stress}
\label{sec:results_stress_nonlin_step}

In Sec.~\ref{sec:results_stress_linear_step}  we argued that a step stress,
$\sigma(t) = \sigma_0
\Theta(t-t_{\rm w})$, of size
$\sigma_0\ll 1$, induces a strain response $\gamma(t)$ which increases over
time, but remains
linear in $\sigma_0$ for at least as long as the linearized constitutive
equations predict
$\gamma(t)\ll 1$. This is because
$\gamma(t)$ provides an upper bound on the local strain of each element.
Although sufficient to
ensure linearity, this is not always necessary; we require only that the
characteristic strain {\em
of those elements which dominate the stress} is small. For $x>2$ (the
Newtonian regime) the
dominant elements have lifetimes $O(1)$ and so the 
response is linear to indefinite
times so long as
$\sigma_0 \ll 1$ (ensuring $\gdot(t) \ll 1$ for all times $t$).
But, whenever $x<2$, the linear
analysis of
Sec.~\ref{sec:results_stress_linear_step} indicates the dominant
elements have lifetimes of
order $t-t_{\rm w}$; so a self-consistently linear response is maintained only
provided that $\gdot(t)(t-t_{\rm w}) \ll 1$, just as in startup of
steady shear (see 
Sec.~\ref{sec:steady_shear_nonlin}; here we make the additional
assumption that $\gdot$ only changes negligibly between $t_{\rm w}$ and $t$).
Using the forms for $J(t-t_{\rm w},t_{\rm w})$ as summarized in
table~\ref{table:compliance}, we then find that
for $1<x<2$ the strain response to step stress remains linear only for as
long as
$t-t_{\rm w} \ll (1/\sigma_0)^{1/(x-1)}$. Beyond this time we expect
strain-induced yielding to become
important.

\begin{figure}[h]
\begin{center}
\epsfig{file=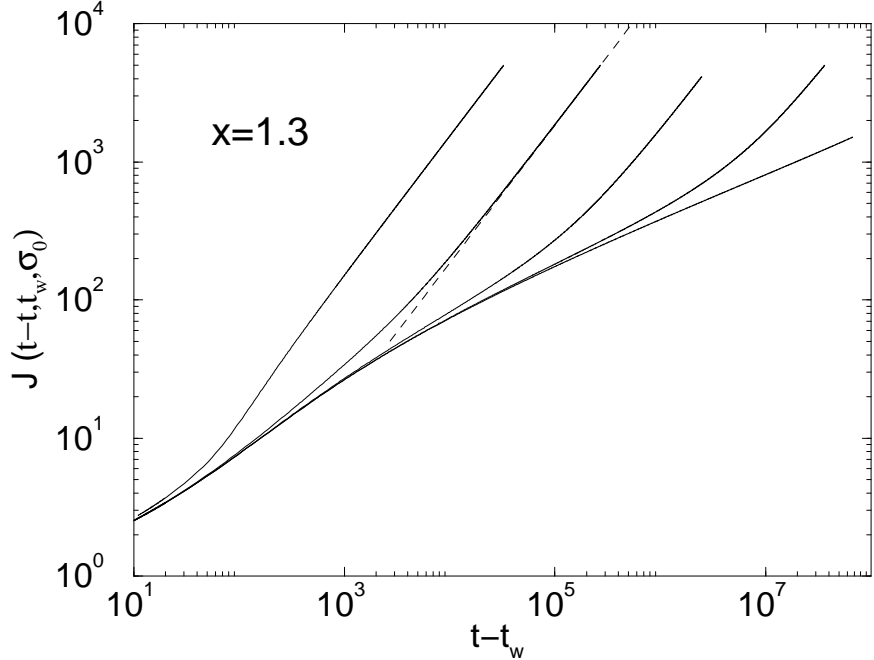,width=11.5cm}
\end{center}
\caption{Nonlinear creep compliance $J(t-t_{\rm w},t_{\rm w},\sigma_0)$
  as a function of time interval $t-t_{\rm w}$, for a step stress
  of size $\sigma_0$
  applied at time $t_{\rm w}=100$. The noise temperature is $x=1.3$.
  Solid lines, bottom to top:
  $\sigma_0=10^{-3}\,,10^{-2.5}\,,10^{-2}\,,10^{-1.5}\,,10^{-1}$. Over
  the time intervals shown, the curve for $\sigma_0 = 10^{-3}$ is
  indistinguishable from the linear compliance (not shown). Dotted
  line: final flow behaviour predicted from steady state flow curve
  for $\sigma_0=10^{-1.5}$.
\label{fig:step_stress_above_glass} }
\end{figure}

To confirm the predicted linearity at short times, and to extract the long
time non-linear
behaviour, we numerically solved the non-linear constitutive
equations~(\ref{eqn:ceone}, \ref{eqn:cetwo}) by an iterative method (see
App.~\ref{app:iterative_code});
this was done first for
$1<x<2$ (Fig.~\ref{fig:step_stress_above_glass}). The results show a
linear regime of the expected
temporal extent, followed by a crossover into a non-linear steady-state
flow regime, in which
$\gamma(t)\propto
\sigma_0^{1/(x-1)}t$. The latter is in agreement with the flow
curve~(\ref{eqn:flow_curve}).

The same numerical procedure was then used for the glass phase, $x<1$, for
which the flow curve
shows a finite yield stress, $\sigma_{\rm y}(x)$. As expected, the
numerical results for step
stress of very small amplitude
$\sigma_0 \ll \sigma_{\rm y}$ show no crossover to a steady flow regime at
late times. Instead, the
system continues to creep logarithmically, according to the linear creep result
(table~\ref{table:compliance}):
\begin{equation}
\gamma(t)= \sigma_0 J(t-t_{\rm w},t_{\rm w}) =
\sigma_0\frac{1}{\psi(1)-\psi(x)}\log\left(\frac{t-t_{\rm w}}{t_{\rm w}}\right)
\end{equation}
The resulting value of $\gdot(t)(t-t_{\rm w})$ never becomes large; so
this is self-consistent.

\begin{figure}[h]
\begin{center}
\epsfig{file=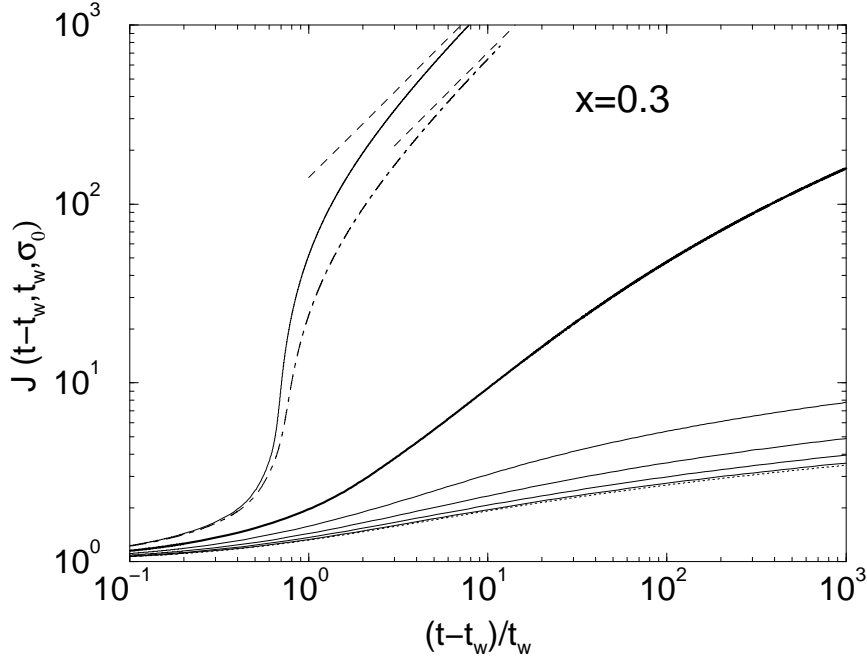,width=11.5cm}
\end{center}
\caption{Nonlinear creep compliance $J(t-t_{\rm w},t_{\rm
  w},\sigma_0)$ as a function of scaled time interval $(t-t_{\rm
  w})/t_{\rm w}$, for a step stress of size $\sigma_0$ applied at time
  $t_{\rm w}$. The noise temperature is $x=0.3$.  Solid curves, bottom
  to top: $\sigma_0/\sigma_{\rm y}=0.2\,,0.4, \ldots,1.2$, all for
  $t_{\rm w}=100$.  The case $\sigma_0=\sigma_{\rm y}$ is shown in
  bold; the dotted curve is the linear response result
  ($\sigma_0\rightarrow 0$).
  The dot-dashed curve shows the effect of decreasing the
  waiting time to $t_{\rm w}=50$, for $\sigma_0/\sigma_{\rm y}=1.2$.
  Comparison of the curves for the two different waiting times for
  this stress value shows that before the crossover into flow,
  the response scales with $(t-t_{\rm w})/t_{\rm w}$; once ergodicity has
  been restored and the system flows, on the other hand, scaling with
  $t-t_{\rm w}$ is recovered. The dashed lines are the predictions for
  final flow behaviour (for the stress above yield) from the steady
  state flow curve.
\label{fig:step_stress_below_glass} }
\end{figure}

Next we studied numerically the case where $\sigma_0$ was not small
but remained less than the yield stress $\sigma_{\rm y}$. For stresses
not too close to the yield stress, we found that the creep was still
logarithmic to a good approximation, but now with a nonlinear
dependence of its amplitude on stress: $\gamma(t) \approx \sigma_0
A(\sigma_0) J(t-t_{\rm w},t_{\rm w})$. The prefactor $A(\sigma_0)$
increases rapidly as $\sigma_0$ approaches the yield stress
$\sigma_{\rm y}$ from below. Very close to the yield stress, the creep
ceases to be logarithmic; $\gamma(t)$ then grows more quickly, but with a
strain rate that still decreases to zero at long times. On the basis
of these observations, we suspect that for a given stress $\sigma_0$
the creep will be logarithmic for short times (where ``short times''
might mean the whole time window which is accessible numerically), but
will gradually deviate from this for longer times. The deviation is
expected to be noticeable sooner for stress values closer to yield. We
attempted to verify this conjecture numerically, but were unable to
access a large enough range of values of $\ln\left((t-t_{\rm
w})/t_{\rm w}\right)$ to do so. Note that, for any
$\sigma_0<\sigma_{\rm y}$, the system ages indefinitely, and there is
no approach to a regime of steady flow.

Finally, as expected from the flow curve, only for stress amplitudes
exceeding the yield stress $\sigma_{\rm y}$ (which of course depends
on $x$) did we see an eventual crossover from logarithmic creep to
steady flow at long times; when that happened, we recovered
numerically the flow-curve result, $\gamma(t)\propto
(\sigma_0-\sigma_{\rm y})^{1/(1-x)} (t-t_{\rm w})$.
Fig.~\ref{fig:step_stress_below_glass} shows examples of our numerical
results that illustrate the various
features of nonlinear creep in the glass phase
mentioned above.\footnote{Note that whereas for most other
  shear scenarios we chose to present glass phase results for a noise
  temperature $x=0.7$, we here chose $x=0.3$. The yield stress is
  larger at this value of $x$, giving us a larger window
  $0<\sigma_0<\sigma_{\rm y}$ over which we see ageing and creep
  uninterrupted by a crossover into flow.}

\section{Conclusion}
\label{sec:conclusion}

In this paper we studied theoretically the role of ageing in the rheology
of soft materials. We first provided, in Sec.~\ref{sec:rheology} a
general formulation of the linear and nonlinear rheological response
functions suited to samples that show ageing, in which time
translation invariance of material properties is lost. (Our analysis
extends and, we hope, clarifies that of~\citet{Struik78}.) This was
followed in Sec.~\ref{sec:ageing} by a review of the concept of
ageing, formally defined by the presence of long term memory, which
can be either weak or strong. We suggested that for many rheological
applications the main interest is in systems with weak long term
memory: these have properties that are age-dependent, but not
influenced by perturbations of finite duration that occurred in the
distant past. We conjectured that weak long term memory is sufficient
to cause the age-dependent linear viscoelastic modulus to become
independent of the start time $t_{\rm s}$ of the oscillatory shear
($G^*(\omega,t,t_{\rm s}) \to G^*(\omega,t)$) while retaining a dependence
on system age $t$; for it to then obey the usual Fourier relation with
the linear step strain response (likewise dependent on age $t_{\rm w}$); and
for it to obey a reciprocal relation $G^*(\omega,t)J^*(\omega,t) = 1$
with the time-varying compliance, similarly defined. Pending a general
proof of these conjectures, all such relationships between
age-dependent rheological quantities do however require empirical
verification for each experimental system, or theoretical model, that
one studies.

Within this conceptual framework, we then explored rheological ageing
effects in detail for the SGR model. After reviewing the basic
rheological definition of the model in Sec.~\ref{sec:model}, we
discussed in Sec.~\ref{sec:ageing_sgr} its ageing properties from the
point of view of the mean jump rate $Y(t)$ whose behaviour is
radically different in the glass phase (noise temperature $x<1$) from
that in the normal phase ($x>1$). The glass phase of the SGR model is
characterized by ``weak ergodicity breaking", which means that the
elastic elements that it describes evolve forever towards higher yield
thresholds (deeper traps), causing a progression toward more elastic
and less lossy behaviour.  Within the glass phase, there is a yield
stress $\sigma_{\rm y}$, and for applied stresses less than this, genuine
ageing effects arise. These phemonena were explored in depth in
Sec.~\ref{sec:results_strain} and Sec.~\ref{sec:results_stress} for
the cases of imposed stress and imposed strain respectively. Ageing
effects are distinguished from otherwise similar transient phenomena
(arising, for example, when $x>1$) by the criterion that a significant
part of the stress relaxation, following infinitesimal step strain,
occurs on timescales that diverge with the age of the system at the
time of strain application. This rheological definition appears
appropriate for most soft materials and follows closely the definition
of long-term memory in other areas of
physics~\cite{CugKur95,BouCugKurMez98,CugKur93}.

In the glass phase of the SGR model, the nature of the ageing is
relatively simple; for a step strain or stress applied at time $t_{\rm w}$,
both the linear stress relaxation function $G(t-t_{\rm w},t)$ and the linear
creep compliance $J(t-t_{\rm w},t_{\rm w})$ become
functions of the scaled time
interval $(t-t_{\rm w})/t_{\rm w}$ only. This scaling is a simple example
of the `time waiting-time superposition' principle postulated
empirically by~\citet{Struik78} (in the somewhat different
context of glassy polymers).
The time-dependent viscoelastic spectra
$G'(\omega,t)$ and $G''(\omega,t)$ have the characteristic ageing
behaviour shown in Fig.~\ref{fig:possible_modes}: a loss modulus that
rises as frequency is {\em lowered}, but falls with age $t$, in such a
way that it always remains less than $G'(\omega,t)$ (which is almost
constant by comparison).  For $x<1$ such spectra collapse to a single
curve (see Fig.~\ref{fig:G_osc_time_dep}) if $\omega t$, rather than $\omega$,
is used as the independent variable.  Note that in more complicated
systems, Eq.~\(eqn:general_ageing) may be required instead, to
describe ageing on various timescales that show different divergences
with the sample age $t_{\rm w}$.  Even in simple materials, there may be an
additional non-ageing contribution to the stress relaxation which the
SGR model does not have; this will also interfere with the scaling
collapse of both $G(t-t_{\rm w},t_{\rm w})$
 and $G^*(\omega,t)$. We found that, in
its glass phase, the SGR model has weak long term memory, and we
confirmed numerically that the conjectured relationships,
Eqs.~(\ref{eqn:secondgstar},\ref{eqn:maybe},\ref{eqn:maybe2}), among
age-dependent linear rheological quantities indeed hold in this case.

Significant ageing was also found for nonlinear rheological responses
of the SGR model. For example the nonlinear step-strain relaxation
follows the same ageing scenario as the linear one, except that all
relaxation rates are speeded up by a single strain-dependent factor
(Eq.~\(eqn:nonlin_step_final)). This form of nonlinearity is a
characteristic simplification of the SGR model, and would break down
if the elastic elements in the model were not perfectly Hookean
between yield events.  Another interesting case was startup of steady
shear; here there is no significant ageing in either the initial
(elastic) or the ultimate (steady flow) regime; yet, as shown in
Fig.~\ref{fig:startup}, the intermediate region shows an overshoot
that is strongly dependent on sample age.  For an old sample, the
elastic elements have higher yield thresholds.
The linear elastic regime therefore extends
further before the imposed strain finally causes yielding, followed by
a larger drop onto the same steady-shear plateau. The plateau itself
is age-independent: the presence of a finite steady flow rate, but not
a finite stress, is always enough to interrupt the ageing process
within the SGR model.  Finally we found that the nonlinear creep
compliance (Fig.~\ref{fig:step_stress_below_glass}), shows interesting
dependence on both the stress level and the age of the sample; for
small stresses we found logarithmic creep (for all $x<1$), crossing
over, as the yield stress is approached, to a more rapid creep that
nonetheless appears to have zero strain rate in the long time limit.
Nonlinear creep gives challenging computational problems in the SGR
model, which is otherwise simple enough, as we have shown, that almost
all its properties can be calculated either by direct asymptotic
analysis or using (relatively) standard numerics. Remaining drawbacks
include (from a phenomenological viewpoint) the lack of tensorial
elasticity in the model and (from a fundamental one) uncertainty as to
the proper physical interpretation, if one indeed exists, of the noise
temperature $x$~\cite{SolLeqHebCat97, long_El}.

Though obviously oversimplified, the SGR model as explored in this paper
may provide a valuable paradigm for the experimental and theoretical study
of rheological ageing phenomena in soft solids. More generally, the
conceptual framework we have presented, which closely follows that
developed to study ageing in non-flowing systems such as spin-glasses,
should facilitate a quantitative analysis of rheological ageing phenomena
across a wide range of soft materials.
 
\appendix

\section{Calculation of Linear Response Properties}

\subsection{Initial Condition}

In discussing the SGR model's non-equilibrium behaviour
(Secs.~\ref{sec:results_strain} and~\ref{sec:results_stress}) we
considered for definiteness a system prepared by a quench from an
infinite noise temperature (see Sec.~\ref{sec:sample_preparation}),
{\em i.e.}, with an initial distribution $P_0(E)=\rho(E)$ of yield
energies or trap depths. For our predictions to be easily compared to
experimental data, however, they must be largely independent of the
details of sample preparation. To test for such independence, we
consider the extent to which our results would change if the
pre-quench temperature, which we denote by $x_0$, were finite.
This corresponds to an initial trap depth
distribution
\begin{equation}
P_0(E) \propto \exp(E/x_0)\rho(E)
\label{eqn:PnE}
\end{equation}
In this appendix, we restrict ourselves to the linear response regime,
where the effects of finite $x_0$ (if any) are expected to be most
pronounced; nonlinearity tends to eliminate memory effects. The same
is true for high temperatures, and correspondingly we will find that
the influence of $x_0$ on our results is confined mainly to final
(post-quench) temperatures $x$ within the glass phase ($x<1$).

\subsection{Yield Rate}
\label{app:yield_rate}

The yield or hopping rate is the basic quantity from which other
linear response properties can be derived; see
eqs.~(\ref{eqn:lin_step},\ref{eqn:G_osc_lin},\ref{eqn:con_J}). It can
be calculated from the second constitutive equation~\(eqn:cetwo)
\begin{equation}
1=G_0(t) + \int_0^t Y(t')G_{\rho}(t-t')dt'
\label{eqn:cetwo_app}
\end{equation}
where we have replaced $Z(t,t')$ by $t-t'$, as is appropriate in the
linear response regime. The function $G_0(t)$ is defined
in~\(eqn:G_rho); for the initial condition~\(eqn:PnE) it is related to
$G_\rho$ via
\[
G_0(t)=G_{\rho}(t,y), \qquad y=x(1-1/x_0)
\]
where we have now included explicitly the noise temperature argument
$(y)$ in the argument list of $G_{\rho}$. Substituting this
into~\(eqn:cetwo_app), and taking Laplace transforms with $\la$ as our
reciprocal time variable, we get:
\begin{equation}
\frac{1}{\la}=\bar{G}_{\rho}(\la,y) + \bar{Y}(\la)\bar{G}_{\rho}(\la,x)
\end{equation}
and hence
\begin{equation}
\label{eqn:Y_transform}
\bar{Y}(\la)=\frac{\frac{1}{\la}-\bar{G}_{\rho}(\la,y)}
{\bar{G}_{\rho}(\la,x)}
\end{equation}
in which (taking Laplace transforms of~\(eqn:G_rho))
\begin{equation}
\label{eqn:G_transform}
\bar{G}_{\rho}(\la,x)=x\int_1^\infty\frac{\tau^{-x-1}}{\la+\tau^{-1}}\,d\tau
=x\int_1^\infty\frac{\tau^{-x}}{1+\la \tau}\,d\tau
\end{equation}
In its present form~\(eqn:Y_transform) cannot be inverted
analytically. We will focus on the long time regime, however, where
progress can be made by using an alternative expression for
$\bar{G}_\rho$. From~\(eqn:G_transform), $\bar{G}_\rho(\la,x)$ has
poles at $\la=-\tau^{-1}$. Because of the integration over all
$\tau=1\ldots\infty$, these poles combine into a branch cut
singularity on the (negative) real axis between $\la=-1$ and
$\la=0$. We will now derive an expression for $\bar{G}_\rho$ that is
valid near this branch cut. This expression does introduce spurious
singularities on the negative real axis for $\la<1$. But after
inversion of the Laplace transform these only give contributions to
$G_\rho(t)$ decaying at least as fast as $\exp(-t)$; they can
therefore be ignored in the long-time limit. We first
write~\(eqn:G_transform) as
\begin{equation}
\frac{1}{x}\,\bar{G}_{\rho}(\la,x)=
\int_0^\infty\frac{\tau^{-x}}{1+\la \tau}\,d\tau -
\int_0^1     \frac{\tau^{-x}}{1+\la \tau}\,d\tau
\label{eqn:integral_split}
\end{equation}
After the rescaling $\la\tau\to\tau$, the first term becomes a
representation of the Beta function.\footnote{The rescaling can be
  carried out only when $\lambda$ is real and positive. But by
  analytic continuation, the result~(\protect\ref{eqn:G_transform_b})
  also holds for complex $\lambda$
  outside the branch cut of $\la^{x-1}$, {\em i.e.}, everywhere except
  on the negative real axis.} In the second term, because now
$\tau\leq1$, we can expand the denominator into a series that is
convergent for $|\la|<1$.  This gives the desired expression
\begin{equation}
\label{eqn:G_transform_b}
\bar{G}_{\rho}(\la,x)=a(x)\la^{x-1}+\sum_{n=0}^{\infty}b_n(x)\la^n
\end{equation}
in which
\begin{equation}
\label{eqn:a} a(x)=x\Gamma(x)\Gamma(1-x), \qquad
\label{eqn:b} b_n(x)=\frac{x(-1)^{n+1}}{n+1-x}
\end{equation}
This is valid for $|\la|<1$ and therefore in particular near the
branch cut $\la=-1\ldots 0$; in the
representation~\(eqn:G_transform_b), this branch cut is apparent in
the fractional power of $\la$ in the first term.  The above derivation
applies a priori only for $x<1$, because otherwise the integrals
in~\(eqn:integral_split) diverge at the lower end. However, using the
relation
\[
\frac{1}{x+1}\,\bar{G}_\rho(\la,x+1)=
\frac{1}{x}-\frac{\la}{x}\,\bar{G}_\rho(\la,x)
\]
which follows directly from~\(eqn:G_transform), it can easily be shown
that~\(eqn:G_transform_b) holds for all $x$. (For integer $x$, there
are separate singularities in the first and second term
of~\(eqn:G_transform_b), but these just cancel each other.)

We can now substitute~(\ref{eqn:G_transform_b},\ref{eqn:a})
into~\(eqn:Y_transform)
and expand the denominator to find a readily invertible expression for
$\bar{Y}(\la)$. Clearly the 
manner in which we perform the expansion depends on whether $x>1$ or
$x<1$. Abbreviating $a(x)=a$, $a(y)=a'$, and $b_n(x)=b_n$, we have
for $x>1$:
\[
\bar{Y}(\la)=\frac{1}{\la}
\left[\frac{1}{b_0} - \frac{a}{b_0^2}\la^{x-1} - \frac{a'}{b_0}\la^{y}
+ O\left(\la^{2(x-1)},\la^{y+x-1},\la, \ldots\right)\right]\\
\]
which, upon inversion of the Laplace transform, gives:
\be
\label{eqn:Y_above_glass} Y(t)=\frac{1}{b_0}-
     \frac{1}{\Gamma(2-x)}\frac{a}{b_0^2}t^{1-x}-
     \frac{1}{\Gamma(1-y)}\frac{a'}{b_0}t^{-y}+
     O\left(t^{2(1-x)},t^{1-x-y},\ldots\right)
\ee
the first term of which is the asymptotic expression for $Y(t)$ above
the glass points, as in~\(eqn:hopping_rate).  For $x<1$ on the other
hand, we have
\be
\label{eqn:Y_transform_below_glass}
\bar{Y}(\la) = \frac{1}{\la}\left[\frac{\la^{1-x}}{a}-
                \frac{b_0\la^{2(1-x)}}{a^2}-\frac{a'}{a}\la^{y+1-x}+
                O\left(\la^{3(1-x)},\la^{y+2(1-x)}, \ldots\right)\right]
\ee
which can be inverted to give
\be
\label{eqn:Y_below_glass} Y(t)=\frac{1}{\Gamma(x)}\frac{t^{x-1}}{a}-
     \frac{1}{\Gamma(1+2(x-1))}\frac{b_0 t^{2(x-1)}}{a^2}
-\frac{a'}{a\Gamma(x-y)}t^{x-1-y}+O\left(t^{3(x-1)},t^{2(x-1)-y}\right)
\ee
the first term again being in agreement with~\(eqn:hopping_rate).
Finally, to obtain $Y(t)$ {\em at} the glass point $x=1$ we
rewrite~\(eqn:Y_transform_below_glass) as:
\[
\bar{Y}(\la)=-\frac{1}{b_0\la}\left[\sum_{n=1}^{p}z^n(\la)
              +O\left(\la^y, \la, \la^{(p+1)(1-x)}\ldots\right)\right]
\]
in which $z(\la)=-b_0\la^{1-x}/a$ and $p$ is the largest integer which
is less than $1/(1-x)$. Inversion of the Laplace transform gives
\[
Y(t)=\frac{-1}{b_0}\sum_{n=1}^{p}\frac{z^n(t)}{\Gamma(1+n(x-1))} +
     O\left(t^{-y}, t^{(p+1)(x-1)}, \ldots\right)
\]
in which $z(t)=-b_0t^{x-1}/a$. The Gamma function can now be expanded
around $\de=1-x=0$; the sum over $p$ can be performed explicitly for
each term in this expansion. Retaining only the dominant terms for
small $\de$, and also taking the limit $\de\to 0$ of the quantities
$z(t)$, $a$ and $b_0$, one finds eventually
\[
\lim_{x\rightarrow
1}Y(t)=\frac{1}{\ln(t)}+\frac{\Gamma'(1)}{\ln^2(t)}+O\left(\frac{1}{\ln^3(t)}
\right)
\]
as stated in~\(eqn:hopping_rate).

Consider now the effect of the pre-quench temperature $x_0$ on the
above results for the asymptotic behaviour of the hopping rate
$Y(t)$. We note first that all the leading terms are independent of
$y$ and hence of $x_0$. For $x>1$, the largest $y$-dependent
subleading term ($t^{-y}$) in~\(eqn:Y_above_glass) becomes more
important for smaller pre-quench temperatures $x_0$.  However,
provided we restrict ourselves to the regime $x_0>x$ ({\it i.e.}, to a
non-equilibrium situation in which a quench is actually performed;
$x=x_0$ corresponds to equilibrium conditions), we see that $y>x-1$
and that, even to subleading order, $Y(t)$ is independent of
$x_0$. (We note furthermore that in the case of the deep quench
defined in Sec~\ref{sec:sample_preparation}, $y=x$ and the term
$t^{-y}$ is very small.) For $x<1$, in~\(eqn:Y_below_glass), the
relative importance of the largest $y$-dependent term ($t^{x-1-y}$)
again depends upon the relative values of the pre- and post-quench
temperatures. For a high enough pre-quench temperature (specifically,
provided $y>1-x$, {\it i.e.}, provided $x_0>x/(2x-1)$) the leading and
subleading terms of $Y(t)$ are independent of $x_0$.  For any
post-quench temperature $x<1/2$, the subleading term necessarily
depends upon $x_0$ since the condition defined above for independence
cannot be satisfied. (This is physically intuitively reasonable, since
in general we expect a system at a lower temperature to remember its
initial condition more strongly.)

\subsection{Step Strain and Oscillatory Strain Response}
\label{app:step_strain}
\label{app:osc_strain}

Once the yield rate $Y(t)$ is know, the linear stress response
$G(t-t_{\rm w},t_{\rm w})$ to a step strain can be calculated
from~\(eqn:lin_step). To get its asymptotic behaviour for $t-t_{\rm
w}\gg1$, $t_{\rm w}\gg 1$, the two regimes in which the time interval
$t-t_{\rm w}$ is much less and much greater than the age at the time
of stress application $t_{\rm w}$ have to be considered separately. In
the first regime ($t-t_{\rm w}\ll t_{\rm w}$), one can Taylor expand
the hopping rate $Y$ around its value at time $t$. In the second
regime, we rewrite~\(eqn:lin_step) as
\be
G(t-t_{\rm w},t_{\rm w})=G_0(t)+\int_0^{t_{\rm w}}Y(t')G_{\rho}(t-t')dt'
\label{eqn:Gstep_long_time_basic}
\ee 
The first term on the right-hand side can then be shown to be
subdominant (at least for $x_0\to\infty$; see
Sec.~\ref{app:finite_temp} below), and the second can be treated by
expanding $G_\rho(t-t')$ around $t'=0$. To leading order, one then
finds the results in table~\ref{table:modulus}. The asymptotic
behaviour of the stress response to oscillatory strain,
$G^*(\omega,t,t_{\rm s})$, is obtained in a similar manner
from~\(eqn:G_osc_lin).

\subsection{Rheological Irrelevance of Initial Condition}

\label{app:finite_temp}

In App.~\ref{app:yield_rate} we discussed the influence of the initial
state of the sample, as parameterized by the ``pre-quench''
temperature $x_0$, on the yield rate $Y(t)$. Now we consider the
effects of $x_0$ on the various (linear) rheological observables,
concentrating on the regime $x<1$ where such effects are expected to be most
pronounced. We begin with the response to a step strain, $G(t-t_{\rm
  w},t_{\rm w})$. In the short time regime $t-t_{\rm w}\ll t_{\rm w}$,
it follows directly from~\(eqn:lin_step) that $x_0$ affects only
subdominant terms (through its effect on $Y(t)$). In the long time
regime $t-t_{\rm w}\gg t_{\rm w}$, we see similarly
from~(\ref{eqn:Gstep_long_time_basic}) that any effect on the leading
behaviour can only be through the first term on the right-hand side,
$G_0(t)=G_\rho(t,y)\sim t^{-y}$. Comparing this with the second term,
which from table~\ref{table:modulus} is $\sim(t_{\rm w}/t)^x$ (note
that $t\approx t-t_{\rm w}$ in the long time regime), and using
$y=x(1-1/x_0)$, one finds that the effect of $x_0$ is negligible up to
$t\approx t_{\rm w}^{x_0}$. For larger $t$, $G(t-t_{\rm w},t_{\rm w})
\approx G_0(t)\approx G_0(t-t_{\rm w})$ and the response is TTI to
leading order. An intuitive explanation for this behaviour can be
found by analysing the evolution of the relaxation time distribution
$P(\tau,t_{\rm w})$ with $t_{\rm w}$~\cite{thesis}. It can be shown
that the initial condition $P(\tau,0)$ is remembered in the long time
tail of this distribution, $\tau\gg t_{\rm w}^{x_0}$. For times $t\gg
t_{\rm w}^{x_0}$, these long relaxation times dominate the behaviour
of $G(t-t_{\rm w},t_{\rm w})$ and cause the observed $x_0$-dependence.

For the step stress response $J(t-t_{\rm w},t_{\rm w})$, we found in
Sec.~\ref{sec:results_stress_linear_step} that memory effects are
rather weaker than for the step strain response. This is because $J$
is sensitive to the average behaviour of the relaxation time
distribution $P(\tau,t')$ over the time interval $t'=t_{\rm w}\ldots
t$, while $G$ depends on $P(\tau,t_{\rm w})$ only. Correspondingly, we
also find that $J(t-t_{\rm w},t_{\rm w})$ is affected only weakly by
the initial preparation of the system and hence by $x_0$. All effects
are in subdominant terms; for the long time behaviour in the glass
phase, for example, one finds that the asymptotic behaviour
$J(t-t_{\rm w},t_{\rm w})\sim\ln((t-t_{\rm w})/t_{\rm w})$ is only
changed by an $x_0$-dependent constant offset~\cite{thesis}.

Finally, consider the oscillatory response functions
$G^*(\omega,t,t_{\rm s})$ and $J(\omega,t,t_{\rm s})$. Any linear oscillatory
perturbation effectively probes only those traps which have a
relaxation time $\tau<1/\omega$. Provided such traps have attained an
$x_0$-independent distribution by the time the perturbation is
switched on at $t_{\rm s}$, $G^*(\omega,t,t_{\rm s})$
and $J^*(\omega,t,t_{\rm s})$ will
be insensitive to $x_0$. It can be shown~\cite{thesis} that the
requirement for this is $\tau\ll t_{\rm s}^{x_0}$ for all $\tau<1/\omega$
and hence $\omega t_{\rm s}^{x_0}\gg 1$. We argue in
App.~\ref{app:ts_indep}, however, that in order to get a sensible
measurement of $G^*$ (and $J^*$) which is independent of {\em start
time} $t_{\rm s}$, we must ensure $\omega t_{\rm s}\gg 1$. This condition then
automatically guarantees that the results are independent of $x_0$.

In summary, the only significant effects of the initial sample
preparation appear in the step strain response at long times ($t\gg
t_{\rm w}^{x_0}$). In the other linear response properties that we
studied, the initial condition only affects subdominant terms. We
reiterate our earlier statement that for nonlinear response, the
initial sample condition should be even less important, because
nonlinearities tend to wipe out memory effects.

\section{Irrelevance of Switch-on Time in the Glass Phase}
\label{app:ts_indep}

It was stated in Sec.~\ref{sec:results_strain_linear_osc} that
$G^*(\omega,t,t_{\rm s})$ does not depend on
$t_{\rm s}$ so long as $\omega(t-t_{\rm s}) \gg 1$ and $\omega t_{\rm
s} \gg 1$. These 
criteria do not depend on the
noise temperature $x$, and therefore hold even in the glass phase, $x\le
1$, where ageing occurs.

This behaviour can be understood as follows. Consider a material which has
not been strained since
preparation except during a time window of duration $t^*$ before the
present time $t$. First write
the linearized constitutive equation as:
\begin{equation}
\label{eqn:rewrite_ceone}
\sigma(t)=-\int_{t-t^*}^t  \gamma(t')\, \frac{dG(t-t',t')}{dt'}\,dt'
\label{eqn:stressage}
\end{equation}
where, for the SGR model
\begin{equation}
\label{eqn:blip_response}
\frac{dG(t-t',t')}{dt'}=-\delta(t-t') +Y(t')G_{\rho}(t-t')
\end{equation} with\footnote{This result follows by differentiation
of~\(eqn:lin_step), respecting
the fact that
$G(t-t_{\rm w},t_{\rm w})$ vanishes for negative $t-t_{\rm w}$
(that is, it contains a factor
$\Theta(t-t_{\rm w})$ which is conventionally suppressed).}
$G_{\rho}(t-t') \sim (t-t')^{-x}$.

Now consider the case of a step strain imposed at $t-t^*$, so that
$\gamma(t)$ is constant in~\(eqn:stressage).  Because $dG/dt'$ contains a
contribution of order
$(t-t')^{-x}$, the integral has significant contributions from $t'$
near $t-t^*$ whenever $x \leq 1$;
in fact in the absence of the factor $Y(t')$ the
integral would not even converge to a finite limit as $t^*$ becomes large.
This is a
signature of long-term memory: Even the strain history in the distant
past has an effect on the stress at time $t$.
On the other hand, for an oscillatory strain
(likewise switched on at
$t-t^*$) one has~\(eqn:stressage) with $\gamma(t) = \gamma_0 e^{i\omega
t}$, and even without the factor $Y(t')$ the integral would now
converge to a finite limit so long as $\omega t^*$ is large. The
convergence of the oscillatory
integral follows from the mathematical result known as Jordan's
lemma~\cite{Copson62} which,
crudely speaking, states that inserting the oscillatory factor $e^{i\omega
t'}$ has a similar effect
to converting the integrand,
$dG/dt'$, to
$\omega d^2G/dt'^2$. (Physically, this extra time derivative arises since
the stress at $t$ due to
any previously executed strain cycle must involve the change in
$dG/dt'$ over the cycle: if this change is small,
the response to positive and negative
strains will cancel.)
Accounting for this extra time derivative, it is simple to check that the
most recently executed
strain cycles indeed dominate the response at time $t$, in contrast to the
non-oscillatory case
where the entire strain history contributes.

This observation allows us to simplify~\(eqn:rewrite_ceone) further by
setting $G(\Delta t,t') \to
G(\Delta t,t)$ where $\Delta t = t-t'$ (we assume $\omega t
\gg 1$, so that the variation in the stress response function over a fixed
number of recent cycles
is negligible). Likewise, the limit of integration can safely be set to
$\Delta=\infty$. Thus we have
\begin{equation}
\label{eqn:G_star_is_ft_one} G^*(\omega,t)=\int_{0}^{\infty} e^{-i
\omega\Delta t}\, \frac{dG(\Delta t,t)}{d\Delta t} \, d\Delta t
\end{equation}
This can be integrated by parts to
give~\(eqn:G_star_is_ft_two) as required.

\section{Numerical Methods}

\subsection{Yield Rate in the Linear Regime}
\label{app:yield_rate_code}

To obtain numerical results for the linear response properties of the
SGR model, the yield rate $Y(t)$ has to be calculated first. A
convenient starting point for this can be obtained by
differentiating~(\ref{eqn:cetwo_app}):
\begin{equation}
Y(t) = -G'_0(t) - \int_0^{t}Y(t') G'_{\rho}(t-t')\, dt'
\end{equation}
This is a Volterra integral equation of the second kind, which can in
principle be solved by standard numerical
algorithms~\cite{PreTeuVetFla92}. Such algorithms are based on
discretizing the time domain into a grid $t_0=0, t_1 \ldots t_n$; the
values $Y_n=Y(t_n)$ are then calculated successively, starting from
the known value of $Y_0$. The subtlety in our case is the choice of
the grid: Because for times $t\gg1$ we expect the hopping rate to be a
power law, we expect relative discretization errors given a time-step
$\Delta t$ to scale roughly as $\Delta t/t_n$.  Once we have chosen an
acceptable (constant) value for the discretization error we are
therefore at liberty to increase the time-step $\Delta t$ linearly
with the time $t_n$, which corresponds to using a geometric time
grid. This allows us to generate data over many decades without too
much computational effort. To improve accuracy, we also used a spline
interpolation between the known points $(t_0, Y_0), (t_1, Y_1) \ldots
(t_{n-1}, Y_{n-1})$ when determining the next value $Y_{n}$.

\subsection{Strain Response to Finite Step Stress}
\label{app:iterative_code}

The numerical scheme used to solve~\(eqn:ceone) and~\(eqn:cetwo) in
the case of an imposed step stress $\sigma(t)=\sigma_0 \Theta(t-t_{\rm w})$
is rather more complicated. This is because both the strain and the
hopping rate, which are coupled through nonlinear integral equations,
have to be calculated as functions of time. Again, we discretize time
into a grid $t_0, t_1\ldots t_n$, where $t_0=t_{\rm w}^+$, and proceed along
the grid calculating the strain $\gamma_n$ and the hopping rate $Y_n$
for successive values of the index $n$.\footnote{Note that the
integral form of the constitutive equations renders the strain and the
hopping rate at any time step $t_n$ dependent upon the values of these
quantities at all previous times -- even times prior to stress
application ($0<t'<t_{\rm w}$). However, for such times the strain is
clearly zero and the hopping rate is identical to that in the linear
response regime, calculated previously.}
The first data point $\gamma_0=\gamma(t_{\rm w}^+)$ and $Y_0=Y(t_{\rm w}^+)$ on
this grid is then obtained directly by treating the discontinuity at
$t_{\rm w}$ ``by hand''. At any subsequent time-step the two non-linear
constitutive equations~\(eqn:ceone) and~\(eqn:cetwo) are solved
simultaneously. The first is essentially of the form:
\begin{equation}
\begin{array}{lclcl}
\label{eqn:effective_ce_one} 0&=&f(\gamma_n, Y_n, \{\gamma_{n'}\},
\{Y_{n'}\}, t_{\rm w}, \sigma_0)
                                            & \mbox{for} & 0\le n'<n
\end{array}
\end{equation}
while the second can be differentiated and rearranged to give
\begin{equation}
\begin{array}{lclcl}
\label{eqn:effective_ce_two} Y_n&=&g(\gamma_n,\{\gamma_{n'}\}, \{Y_{n'}\},
t_{\rm w}, \sigma_0)&
\mbox{for} &0\le n'<n
\end{array}
\end{equation}
Because~\(eqn:effective_ce_one) cannot be solved explicitly for
$\gamma_n$, we use an iterative process. At each time-step we start by
placing sensible upper and lower bounds on $\gamma_n$, derived from
physical expectations about the time dependence of the strain
$\gamma(t)$.  Each bound in turn is substituted
into~\(eqn:effective_ce_two) (to find the corresponding value of
$Y_n$) and (with its $Y_n$) into the function $f$ of the right hand
side of~\(eqn:effective_ce_one). The secant
method~\cite{PreTeuVetFla92} is then used to
update one of the bounds, and the new bound used to calculate a new
$Y_n$ and $f$.  This process is repeated until we obtain a
sufficiently small value of $f$ ($|f|<10^{-8}$). The current values of
$\gamma_n$ and $Y_n$ are then accepted and we proceed to the next
time-step.

We initially chose a geometric grid of time values $t_0, t_1 \ldots$,
but this led to numerical instabilities. We therefore switched to an
adaptive procedure which chooses time-steps such that the strain
always increases by approximately the same amount in a given
time-step.

Finally, note that at each iteration loop of each time-step we in
principle need to evaluate double integrals of the form $I=\int_0^t
h(Z(t,t')) dt'$ in which
\begin{equation}
Z(t,t')=\int_{t'}^t dt'' \exp\{[\gamma(t'')-\gamma(t')]^2/2x\}
\end{equation}
Because this is very costly computationally, we first calculate at
each loop $Z(t,t')$ on a grid of $t'$ values ranging from $0$ to $t$
and set up an interpolation over the calculated points. We are then
left with single integrals of the same form as $I$, and look up
the value of $Z$ whenever the integrand is called.


\end{document}